\documentstyle[psfig,12pt]{article}
\def\refe{\par\noindent\hangindent=1.5cm}
\begin{document}
\begin{center}
{\Large \bf Fermi's Paradox -- The Last Challenge for
Copernicanism?}

\vspace{0.3cm} \large Milan M.\ \'Cirkovi\'c \\
\vspace{0.1cm}
{\it Astronomical Observatory Belgrade, \\
Volgina 7, 11160 Belgrade, Serbia \\

\&

Department of Physics, University of Novi Sad, \\
Trg Dositeja Obradovi\'ca 4, 21000 Novi Sad, Serbia \\
 e-mail: {\tt mcirkovic@aob.bg.ac.rs} }

\begin{abstract}
\noindent We review Fermi's paradox (or the "Great Silence"
problem), not only arguably the oldest and crucial problem for the
Search for ExtraTerrestrial Intelligence (SETI), but also a
conundrum of profound scientific, philosophical and cultural
importance. By a simple analysis of observation selection
effects, the correct resolution of Fermi's paradox is certain to
tell us something about the future of humanity. Already a more
than three quarters of a century old puzzle -- and a quarter of
century since the last major review paper in the field by G.
David Brin -- Fermi's paradox has generated many ingenious
discussions and hypotheses. We analyze the often tacit
methodological assumptions built into various answers to this
puzzle and attempt a new classification of the numerous solutions
proposed in an already huge literature on the subject. Finally,
we consider the ramifications of various classes of hypotheses
for the practical SETI projects. Somewhat paradoxically, it seems
that the class of (neo)catastrophic hypotheses gives, on balance,
the strongest justification for guarded optimism regarding our
current and near-future SETI efforts.
\end{abstract}

\vspace{0.1cm}
\end{center}


\vspace{0.5cm}

\begin{quote}
{\sl If you do not expect the unexpected, you will not find it;
for it is hard to be sought out and difficult.}
\end{quote}
\hspace{4.5cm} Heraclitus of Ephesus (cca.\ 500 BC)

\begin{quote}
{\sl How many kingdoms know us not!}
\end{quote}
\hspace{4.5cm} Blaise Pascal, \textit{Thoughts\/} (cca.~1660)

\begin{quote}
{\sl What's past is prologue...}
\end{quote}
\hspace{4.5cm} William Shakespeare, \textit{The Tempest}, II, 1
(1610-11)

\section{Introduction: Where is Everybody?}
Fermi's paradox (henceforth FP) presents arguably the greatest
challenge for any practical SETI acitivity, as well as one of the
least understood of all "grand questions" posed in the history of
science. As is well known and established by the research of
Jones (1985), the key argument follows a lunchtime remark of the
great physicist, Enriko Fermi: "Where is everybody?" First
discussed in print by the Russian space-science pioneer
Konstantin Eduardovich Tsiolkovsky, and in recent decades
elaborated upon in detail by Viewing, Hart, Tipler and others
(for detailed reviews see Brin 1983, Webb 2002), the argument
presents a formidable challenge for any theoretical framework
assuming a naturalistic origin of life and intelligence. As such,
this should worry not only a small group of SETI enthusiasts, but
challenges some of the deepest philosophical and cultural
foundations of modern civilization. It is hard to conceive a
scientific problem more pregnant in meaning or richer in
connections with the other "big questions" of science throughout
the ages. In addition, it presents a wonderful opportunity for
public outreach, popularization and promotion of astronomy,
evolutionary biology, and related sciences.

Tsiolkovsky, Fermi, Viewing, Hart, and their followers argue on
the basis of two premises:

(i) the absence of extraterrestrials in the Solar System ("Fact A"
of Hart 1975); and

(ii) the fact that they have had, \textit{ceteris paribus}, more
than enough time in the history of Galaxy to visit, either in
person or through their conventional or self-replicating probes.

\noindent Characteristic time for colonization of the Galaxy,
according to these investigators, is what we shall call the
Fermi-Hart timescale (Hart 1975, Tipler 1980):
\begin{equation}
\label{jedan} t_{FH} = 10^6 - 10^8 \; {\rm years,}
\end{equation}
making the fact that the Solar System is (obviously) not
colonized hard to explain, if not for the total absence of
extraterrestrial cultures. It is enough for our purposes to
contend that this timescale is well-defined, albeit not precisely
known due to our ignorance regarding the possibilities and modes
of interstellar travel. For comparison, the accepted age of the
Earth as an object of roughly present-day mass is (All\`{e}gre et
al. 1995)
\begin{equation}
\label{dva} t_\oplus = (4.46 \pm 0.02 ) \times 10^9 \; {\rm
years.}
\end{equation}
The drastic difference between the timescales in (1) and (2) is
one of the ways of formulating Fermi's paradox. In the next
section, we shall see that there is still more serious numerical
discrepancy in play, when we account for the distribution of ages
of terrestrial planets in the Milky Way.

Even more generally, we need not consider the direct physical
contact between an extraterrestrial civilization and Earth or the
Solar System (insofar as we do not perceive evidence of
extraterrestrial visits in the Solar System; however, this is
still an act of faith, considering the volume of space comprising
our planetary system\footnote{In view of this circumstance, it is
occasionally suggested that we also need a Search for
ExtraTerrestrial Artifacts (SETA) programs as well (Freitas and
Valdes 1980, Arkhipov 1996, 1997). Although we neglect this
possibility in the further considerations in this text it worth
noticing that this is a special case of a more generally
understood unorthodox SETI programs which we consider in the
concluding section.}). It is sufficient to consider a weaker
requirement: namely, that no extraterrestrial civilizations are
{\it detectable\/} by any means from Earth at present. This
includes the detectability of astroengineering or macroengineering
projects over interstellar distances (Dyson 1960, Sagan and
Walker 1966, Freitas 1985, Harris 1986, 2002, Zubrin 1995,
Timofeev et al. 2000, Arnold 2005). In the words of the great
writer and philosopher Stanislaw Lem, who authored some of the
deepest thoughts on this topic, Fermi's paradox is equivalent to
the "absence of cosmic miracles" or the {\it Silentium
Universi\/} ("cosmic silence"; Lem 1977, 1984). Following the
classic review by Brin (1983), we may introduce "contact
cross-section" as a measure of the probability of contact -- by
analogy with introduction of cross-sections in atomic and
particle physics -- and reformulate FP as the question why this
cross-section in the Milky Way at present is so small in
comparison to what could be naively expected.

Schematically, Fermi's paradox can be represented as

\vspace{0.2cm}

\noindent {\it spatiotemporal scales of the Galaxy + the absence
of detected extraterrestrial civilizations (+ additional
assumptions) $\rightarrow$ paradoxical conclusion.}

\vspace{0.2cm}

\noindent Here, under spatiotemporal scales we include our
understanding of the age of the Galaxy, the Solar System and the
ages (incompletely known) of other planetary systems in the Milky
Way. The additional assumptions can be further explicated as

\vspace{0.2cm}

\noindent {\it additional assumptions = "naive realism" +
naturalism + Copernicanism + gradualism + non-exclusivity}.

\vspace{0.2cm}

\noindent These assumptions are quite heterogeneous. By "naive
realism" we denote the working philosophy of most of science (as
well as everyday life), implying that there is a material world
out there, composed of objects that occupy space and have
properties such as size, mass, shape, texture, smell, taste and
colour.\footnote {Philosophical literature often calls this view
\textit{direct realism\/} or \textit{common sense realism}.}
These properties are usually perceived correctly and obey the
laws of physics. In the specific case of FP, the basic premise
following from naive realism is that there are, indeed, no traces
of extraterrestrial intelligent presence detected either directly
or indirectly ("Fact A" of Hart 1975). We shall discuss below
some of the hypotheses for resolving FP which directly violate
this realist view; an extreme and ludicrous example -- but
powerfully present in pop-culture -- of such naively anti-realist
standpoint is a view that, contrary to scientific consensus, some
humans are in contact with extraterrestrial visitors and are
conspiring with them (e.g., Barkun 2003). Naive realism and
naturalism (Section 4 below) are methodological assumptions
typically in play in any scientific research. Copernicanism and
gradualism are somewhat more specific tenets, stemming more from
our experiences in the history of physical science than from the
general epistemology. Copernicanism (often called the Principle
of Mediocrity) in a narrow sense tells us that there is nothing
special about the Earth or the Solar System or our Galaxy within
large sets of similar objects throughout the universe. In a
somewhat broader sense, it indicates that there is nothing
particularly special about us as observers: our temporal or
spatial location, or our location in other abstract spaces of
physical, chemical, biological, etc., parameters are typical or
close to typical.\footnote{Note that this does not mean that our
locations in these spaces are \textit{random}. The latter
statement is obviously wrong, since a random location in
configuration space is practically certain to be in the
intergalactic space, which fills 99.99...\% of the volume of the
universe. This is a long-standing confusion and the reason why
Copernicanism is most fruitfully used in conjuction with some
expression of observational selection effects, usually
misleadingly known as the 'anthropic principle'; for detailed
treatment see Bostrom 2002.} Gradualism, on the other hand, is
often expressed as the motto that "the present is key to the
past" (with corollary that "the past is key to the future"). This
paradigm, emerging from geological science in the 19th century
with the work of Charles Lyell -- and expanding, through Lyell's
most famous pupil, Darwin, into life sciences -- has been subject
of the fierce criticism in the last quarter of a century or so.
We shall return to this issue in Section 7.

Finally, the role of the non-exclusivity (or "hardness" in some of
the literature) assumption needs to be elucidated.
Non-exclusivity (following Brin 1983) is simply a principle of
causal parsimony applied to the set of hypotheses for resolving
FP: we should prefer those hypotheses which involve a smaller
number of local causes. FP is eminently {\bf not} resolved by
postulating that a single old civilization self-destructs in a
nuclear holocaust. FP {\bf is} resolved by hypothesizing that
{\bf all} civilizations self-destruct soon after developing
nuclear weapons, but the major weakness of such a solution is
obvious: it requires many local causes acting independently in
uniform to achieve the desired explanatory end. In other words,
such a solution is exclusive (or "soft"). As long as we have any
choice, we should prefer non-exclusive (or "hard") solutions,
i.e., those which rely on a small number of independent causes.
For instance, the hypothesis, we shall discuss in more detail
below, that a $\gamma$-ray burst can cause mass extinction over a
large portion of the Galaxy and thus arrest evolution toward
advanced technological society, is quite non-exclusive.

\section{Recent Developments}

Fermi's Paradox has become significantly more serious, even
disturbing, of late. This is due to several independent lines of
scientific and technological advance occurring during the last
two decades:

\begin{enumerate}
\item
The discovery of more than 350 extrasolar planets so far, on an
almost weekly basis (for regular updates see {\tt
http://exoplanet.eu/}). Although most of them are "hot Jupiters"
and not suitable for life as we know it (some of their satellites
could still be habitable, however; cf. Williams et al. 1997),
many other exoworlds are reported to be parts of systems with
stable circumstellar habitable zones (Noble et al. 2002, Asghari
et al. 2004, Beaug\'e et al. 2005). It seems that only the
selection effects and the capacities of present-day instruments
stand between us and the discovery of Earth-like extrasolar
planets, envisioned by the new generation of orbital
observatories. In addition, this relative wealth of planets
decisively disproves old cosmogonic hypotheses regarding the
formation of the Solar System as a rare and essentially
non-repeatable occurrence, which have been occasionally used to
support skepticism on issues of extraterrestrial life and
intelligence.

\item
Improved understanding of the details of the chemical and
dynamical structure of the Milky Way and its Galactic Habitable
Zone (GHZ; Gonzalez et al. 2001, Pe\~na-Cabrera and
Durand-Manterola 2004, Gonzalez 2005). In particular, the
important calculations of Lineweaver (2001; Lineweaver, Fenner
and Gibson 2004) show that Earth-like planets began forming more
than 9 Gyr ago, and that their median age is $ \langle t \rangle
= (6.4 \pm 0.7) \times 10^9$ yrs - significantly more than the
age of the Earth. This means that the age difference
\begin{equation}
\label{tri} \langle t \rangle - t_\oplus = (1.9 \pm 0.7) \times
10^9 \; {\rm years},
\end{equation}
is large in comparison with the Fermi-Hart timescale in
(\ref{jedan}). This also means that not only the oldest ones, but
a large majority of habitable planets are much older than Earth.
The significance of this result cannot be overstated, since it
clearly shows that the naive naturalist, gradualist and Copernican
view {\bf must be} {\bf wrong}, since it implies that millions of
planets in the Milky Way are inhabited by Gyr-old
supercivilizations, in clear contrast with observations.

\item
Confirmation of the {\bf rapid\/} origination of life on early
Earth (e.g., Mojzsis et al. 1996); this rapidity, in turn, offers
strong probabilistic support to the idea of many planets in the
Milky Way inhabited by at least the simplest lifeforms
(Lineweaver and Davis 2002).

\item
Discovery of extremophiles and the general resistance of
simple lifeforms to much more severe environmental stresses than
had been thought possible earlier (Cavicchioli 2002). These
include representatives of all three great domains of terrestrial
life ({\it Bacteria}, {\it Archaea}, and {\it Eukarya\/}),
showing that the number and variety of cosmic habitats for life
are probably much larger than conventionally imagined.

\item
Our improved understanding of molecular biology and biochemistry
leading to heightened confidence in the theories of the
naturalistic origin of life or \textit{biogenesis\/} (Lahav et
al. 2001, Ehrenfreund et al. 2002, Bada 2004). The same can be
said, to a lesser degree, for our understanding of the origin of
intelligence  and technological civilization -- which we shall
henceforth label \textit{noogenesis} (e.g., Chernavskii 2000).

\item
Exponential growth of the technological civilization on Earth,
especially manifested through Moore's Law and other advances in
information technologies (see, for instance, Schaller 1997,
Bostrom 2000). This is closely related to the issue of
astroengineering: the energy limitations will soon cease to
constrain human activities, just as memory limitations constrain
our computations less than they once did. We have no reason to
expect the development of technological civilization elsewhere to
avoid this basic trend.

\item
Improved understanding of the feasibility of interstellar travel
in both the classical sense (e.g., Andrews 2003), and in the more
efficient form of sending inscribed matter packages over
interstellar distances (Rose and Wright 2004). The latter result
is particularly important since it shows that, contrary to the
conventional skeptical wisdom, it makes good sense to send
(presumably extremely miniaturized) interstellar probes even if
only for the sake of communication.

\item
Theoretical grounding for various
astroengineering/macroengineering projects (Badescu 1995, Badescu
and Cathcart 2000, 2006, Korycansky et al. 2001, McInnes 2002)
potentially detectable over interstellar distances. Especially
important in this respect is the possible combination of
astroengineering and computation projects of advanced
civilizations, like those envisaged by Sandberg (1999).

\item
Our improved understanding of the extragalactic universe has
brought a wealth of information about other galaxies, many of them
similar to the Milky Way, while not a single civilization of
Kardashev's (1964) Type III has been found, in spite of the huge
volume of space surveyed (Annis 1999b).
\end{enumerate}

Although admittedly uneven and partially conjectural, this list of
advances and developments (entirely unknown at the time of
Tsiolkovsky's and Fermi's original remarks and even Viewing's,
Hart's and Tipler's later re-issues) testifies that Fermi's
paradox is not only still with us more than 75 years after
Tsiolkovsky and more half a century after Fermi, but that it is
more puzzling and disturbing than ever.\footnote{One is tempted
to add another item of a completely different sort to the list:
The empirical fact that we have survived more than sixty years
since the invention of the first true weapon of mass destruction
gives us at least a vague Bayesian argument countering the
ideas---prevailing at the time of Fermi's original lunch---that
technological civilizations tend to destroy themselves as soon as
they discover nuclear power. This is not to contest that the
greatest challenges on the road toward securing a future for
humankind still lie ahead of us; see, e.g., Bostrom and
\'Cirkovi\'c (2008).} In addition, we have witnessed substantial
research leading to a decrease in confidence in Carter's (1983)
so-called "anthropic" argument, the other mainstay of SETI
skepticism (Wilson 1994, Livio 1999, \'Cirkovi\'c et al. 2009).
All this has been accompanied by an increase of public interest
in astrobiology and related issues (Des Marais and Walter 1999,
Ward and Brownlee 2000, 2002, Webb 2002, Grinspoon 2003, Cohen
and Stewart 2002, Dick 2003, Chyba and Hand 2005, Michaud 2007).
The list above shows, parenthetically, that the quite widespread
notion (especially in popular the press) that there is nothing
new or interesting happening in SETI studies is deeply wrong.

In the rest of this review, we survey the already voluminous
literature dealing with Fermi's Paradox, with an eye on the
classification scheme which could help in understanding many
hypotheses posed in this regard. FP is fundamentally intertwined
with so many different disciplines and areas of human knowledge,
that it is difficult to give more than a very brief sketch in the
present format. It should be noted right at the beginning that it
is not entirely surprising that several scientific hypotheses for
resolving FP have been formulated, in a qualitative manner, in the
recreational context of SF art; astrobiology is perhaps uniquely
positioned to exert such influence upon human minds of various
bents. After all, much of the scientific interest in questions of
life beyond Earth in the 20th century was generated by works such
as Herbert G. Wells' \textit{War of the Worlds}, Sir Arthur
Clarke's \textit{2001: Space Odyssey}, or Sir Fred Hoyle's
\textit{The Black Cloud}.

In Fig. 1, we schematically present a version of FP based upon
the scenario of Tipler (1980), using self-replicating, von Neumann
probes which, once launched, use local resources in visited
planetary systems to create copies of themselves. It is clear
that the exponential expansion characteristic of this mode of
colonization leads to the lowest values for the Fermi-Hart
timescales (\ref{jedan}). It is important to understand, however,
that although FP is {\bf aggravated} with von Neumann probes, it
is not really dependent on them. FP would still present a
formidable challenge if at some stage it could be shown that
interstellar von Neumann probes are unfeasible, impractical or
unacceptable for other reasons (possibly due to the danger they
will pose to their creators, as speculated by some authors; see
the "deadly probes" hypothesis in Section 7).

Two further general comments are in order. {\bf (I)} Although it
is clear that philosophical issues are unavoidable in discussing
the question of life and intelligence elsewhere in the universe,
there is a well-delineated part of philosophical baggage which we
shall leave at the entrance. This includes the misleading
insistence on definitional issues. The precise definition of both
life and intelligence in general is impossible at present, as
accepted by almost all biologists and cognitive scientists. This,
however, hardly prevents any of them in their daily research
activities. There is no discernible reason why we should take a
different approach in astrobiology and SETI studies and insist on
a higher level of formal precision in these fields. Intuitive
concepts of life and intelligence are sufficiently developed to
enable fruitful research in these fields, in the same manner as
the intuitive concept of life enables research in the terrestrial
biology and other life sciences; or, even more prominently and
dramatically, the intuitive concept of number has enabled
immensely fruitful research in mathematics for millennia before
the advent of set theory as the axiomatic foundation for modern
mathematics finally enabled completely general and formal
definition of number (by personalities such as Frege, Russell,
G\"{o}del, Turing, Church, Kleene, and Post; e.g., Hatcher 1982,
Penrose 1989). The history of science also teaches us that
formalization of paradigms (including precise definitions) occurs
only at later stages of mature disciplines (Butterfield 1962,
Kragh 1996) and there is no reason to doubt that astrobiology
will conform to the same general picture.

\begin{figure}
\psfig{file=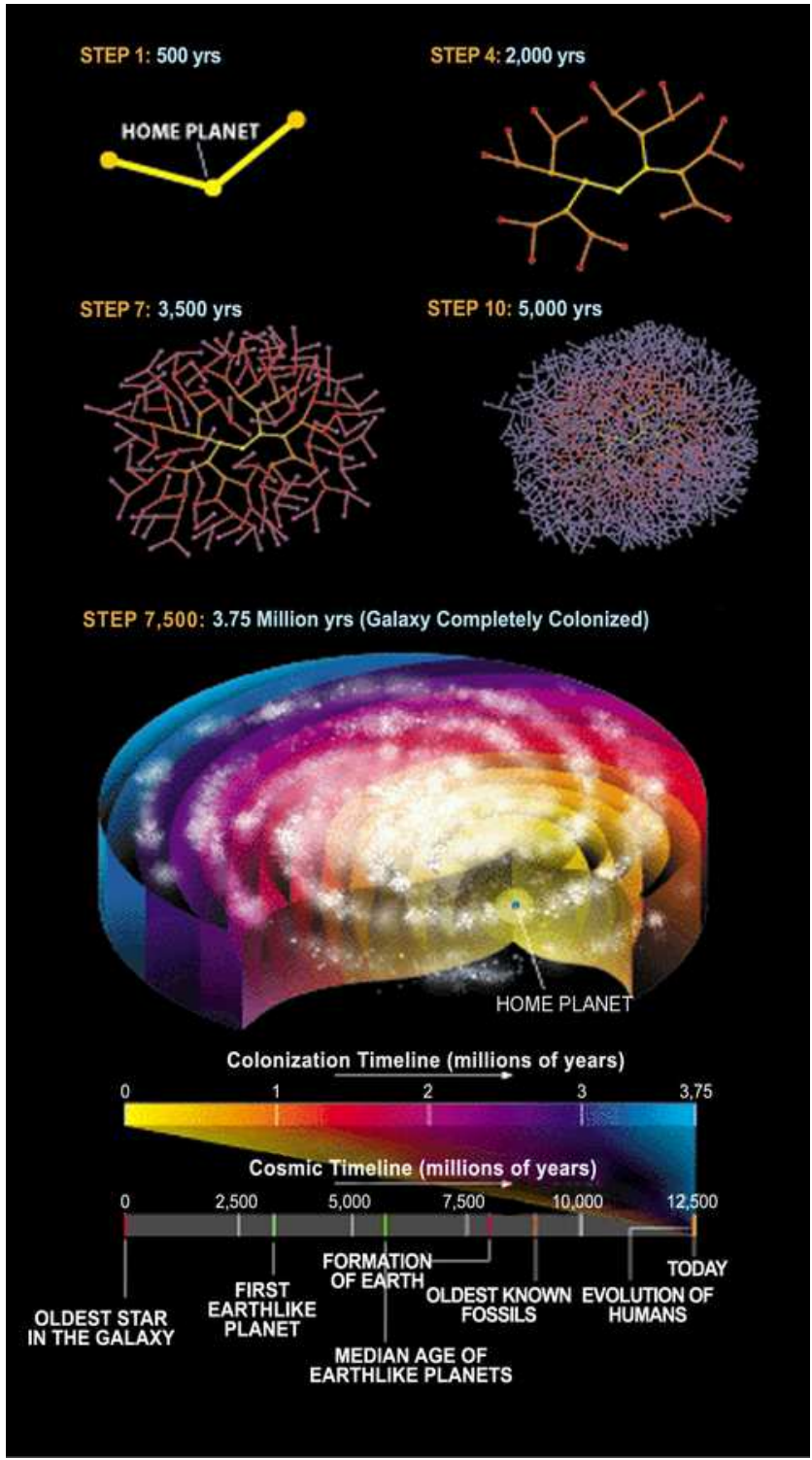,width=11cm} \caption{Fermi's paradox in
a model with slow von Neumann probes, giving a typically low
Fermi-Hart timescale for the colonization of the Milky Way. The
relevant timescales are also shown. \label{pic1}}
\end{figure}

It is clear, for instance, that the Darwinian evolution on Earth
brought about at best a few intelligent species\footnote{The
status of the intelligence of marine mammals is still unclear
(e.g., Browne 2004), while we still do not know whether
undoubtedly intelligent neanderthals were truly separate species,
distinct from \textit{Homo sapiens\/} (e.g., Hawks and Wolpoff
2001).} and only one with technological capacities for engaging
in SETI and similar large-scale cosmic activities. In these
cases, the precise definition of intelligent species (much less a
conscious one; see the disturbing comments of Jaynes 1990 and
Raup 1992, showing that consciousness is in any case much less
significant than colloquially presumed) is unnecessary; while the
awareness that this might be radically different in the SETI
context is desirable, we need to proceed along the same, broadly
operationalist lines. For this reason, we shall use the terms
"extraterrestrial intelligence", "intelligent beings", etc.\ in
their non-technical or vernacular meaning, roughly as
placeholders for beings we are interested in meaningfully
communicating with.

Similarly, we use the locution "advanced technological
civilization" along the lines sketched in \'Cirkovi\'c and
Bradbury (2006), as denoting a community of intelligent beings
capable of manipulating matter and energy on sufficiently large
scale. Again, the precise and useful definition is impossible to
get, but some of the properties of hypothetical advanced
technological civilization would be, for instance, that they are
immune to natural hazards (like those threatening the survival of
humanity at present, like impacts or supervolcanism) and that
consequences of their industrial and computational activities
could, in principle, be detected from interstellar distances. It
is exactly the magnitude of difference in (\ref{tri}) that we
naively expect at least some of the advanced technological
civilization to have arisen in the Milky Way.

{\bf (II)} A useful way of thinking about FP is by analogy with
Olbers' paradox in classical cosmology, as first elucidated by
Alm\'{a}r (1992). Both intentional signals and unintentional
manifestations of advanced technological civilizations in FP are
analogous to the light of distant stars which we would expect, on
the basis of wide spatiotemporal assumptions, to flood us,
terrestrial observers. That this is not happening points to some
flaw in either the reasoning or the assumptions. We now know
(e.g., Wesson et al. 1987) that Olbers' paradox is resolved mainly
by the fact that the stellar population of the universe is of
finite age: the light simply has not had enough time to establish
thermodynamical equilibrium with the cold and empty interstellar
(intergalactic) space. Contrary to a popular opinion --
occasionally found even in astronomy textbooks -- Hubble
expansion actually is actually an almost negligible, minor effect
to take into consideration in resolving Olbers' paradox. FP can,
in principle, also be resolved by the finite age of the stellar
population (and hypothetical extraterrestrial civilizations),
which would correspond to the "rare Earth" class of hypotheses
(see Section 6 below). However, FP is significantly less
constrained and thus allows for additional classes of
explanation, as will be elucidated below. But this analogy
strengthens the general analogy which exists between the current
immature and vigorous stage of astrobiology and the state in
which physical cosmology had been in the 1920s and 1930s (Kragh
1996, 2007, Dick 1996, 2003).

\section{What's Past is Prologue}

It was noticed as early as the Byurakan conference (Sagan 1973)
that the search for extraterrestrial intelligence and the issue
of the future of intelligence here, on Earth, are closely linked.
If we accept Copernicanism, than within reasonable temporal and
physical constrains, we expect the status of evolution on Earth
to reflect the Galactic average for given age of our habitat.
This is exactly the rationale for the assumption (widely used in
the orthodox SETI; e.g., Shklovskii and Sagan 1966, Tarter 2001,
Duric and Field 2003) that most of the members of the
hypothetical "Galactic Club" of communicating civilizations are
significantly older than ours.\footnote{The magnitude of the age
difference has been, however, constantly underestimated, as was
the case even before the results of Lineweaver cited above became
available. The orthodox SETI literature does not discuss the age
differences of the order of Gyr, which is indicative of the
optimistic bias on the part of the authors.} This applies to the
future as well -- the status of extraterrestrial biospheres older
than the Earth reflects, on the average, the {\bf future} status
of the terrestrial biosphere. This reflects a deeper tension at
the very heart of FP: belief in unlimited progress, coupled with
the Copernican assumption, leads to either contradiction or bleak
prospects for our future.

This is especially pertinent and disturbing in view of Fermi's
paradox. The fact that we observe no supercivilizations (of
Kardashev's Type III, for example) in the Milky Way, in spite of
plentiful time for their emergence, is \textit{prima facie\/}
easiest to  explain by postulating the vanishing probability or
impossibility of their existence in general. An obvious
consequence is that, for humanity or its descendants, the
transformation into a supercivilization is either overwhelmingly
unlikely or flatly impossible. But the cut goes deeper both ways
-- if, as some disenchanted SETI pioneers (in particular Iosif
Shklovskii and Sebastian von Hoerner; see, e.g., von Hoerner 1978
and comments in Lem 1977) argued, the reason behind the absence of
extraterrestrial signals is the prevalent self-destruction of
each individual extraterrestrial civilization (for instance,
through nuclear annihilation soon after the discovery of nuclear
energy), this would mean that humanity is also overwhelmingly
likely to self-destruct. If natural hazards (in the form of, for
example, impacts by comets and asteroids or supervolcanic
eruptions; cf. Chyba 1997; Rampino 2002) are the main culprits
beyond the absence of extraterrestrials -- automatically implying
that they are, on average, more frequent than inferred from the
terrestrial history thus far, which might be a consequence of the
anthropic bias (cf. Bostrom 2002, \'Cirkovi\'c 2007) -- then we,
humans, have statistically bleak prospects when faced with similar
natural catastrophes. Consequently, in such case we would have to
ascribe our surviving thus far to sheer luck, which holds no
guarantees for the future. And the same applies to whatever
causative agent causes the contact cross-section to be extremely
small; for instance, if intelligent communities remain bound to
their home planets in a form of cultural and technological stasis
due to imposition of global totalitarianism which, provided
technological means already clearly envisioned (Caplan 2008),
could permanently arrest progress, this would mean that our own
prospects of avoiding such a hellish fate are negligible. In that
sense, the astrobiological history of the Milky Way is a
Shakespearian prologue to the study of the future of humanity.

Exactly this form of "mirroring" of whatever provides the
solution to Fermi's paradox is the reason why some of the
researchers interested in the future of humanity are expressing
their hopes that the Earth is unique in the Galaxy, at least in
terms of evolving intelligent beings (e.g., Hanson 1998a, Bostrom
2008). This would correspond to those solutions of FP rejecting
Copernicanism (see Section 6 below), which these authors consider
a lesser evil. However, such a form of pessimism is not mandatory
-- we can have both optimism toward SETI and optimism about
humanity's future. This forms one of the motivations for
developing some of the neocatastrophic solutions to FP (Section
7) which avoid this tension.

\section{Naturalism and Continuity} \label{natu}

The successes of science since the so-called "Scientific
Revolution" of the 17th century (celebrated, among other things,
in the International Year of Astronomy 2009, as 400 years since
Galileo's invention of the telescope and consequent revolutionary
discoveries) have led to a worldview that could be called
naturalistic, since it assumes the absence of supernatural forces
and influences on the phenomena science is dealing with (Kuhn
1957, Butterfield 1962). Here, as in the case of intelligence, we
are using a rough, non-technical definition which is entirely
sufficient for meaningful discussion.\footnote{It might be
interesting to note that Alfred Russell Wallace, co-discoverer of
natural selection with Darwin, has in several regards been a
precursor to contemporary astrobiology and in particular to the
study of FP. Beside speculating on life on Mars in a separate
treatise, in his fascinating book \textit{Man's Place in the
Universe\/} (Wallace 1903), preceding even Tsiolkovsky's
formulation of FP by about three decades, he argued that
naturalism cannot account for the fine-tuned structure of the
universe. That was perhaps the last attempt at large-scale denial
of naturalism in what can be regarded as a legitimate scientific
context.}

One of the central issues of astrobiology is to what extent we
can talk about biogenesis (and, by extension, noogenesis) in
naturalistic terms. This issue has been investigated in depth by
Fry (1995, 2000), who showed that a necessary ingredient in any
scientific account of biogenesis is the so-called {\bf continuity
thesis}: "the assumption that there is no unbridgeable gap between
inorganic matter and living systems, and that under suitable
physical conditions the emergence of life is highly probable."
Adherence to the continuity thesis, as Fry demonstrates, is a
precondition for scientific study of the origin of life;
contrariwise, the views that biogenesis is a "happy accident" or
"almost miracle" are essentially creationist, i.e., unscientific.
The classification suggested below relies on this analysis of the
continuity thesis and in part on its extension to
noogenesis.\footnote{Whether such an extension is legitimate,
remains an open question, one too difficult to be tackled here. We
mention in passing that at least one of the proposed solutions
discussed below -- the adaptationist hypothesis of Raup (1992) and
Schroeder (2002) -- explicitly denies this generalization.}

The continuity thesis has been supported by many distinguished
scientist throughout history, but none did more to promote it
than the great British polymath John B.\ S.\ Haldane (1892-1964).
In both his research writings in biology, mathematics, astronomy,
etc., and in philosophical essays (especially Haldane 1972
[1927]), he insisted on the continuity between physical (in
particular cosmological), chemical, biological and even cultural
evolution. Haldane was a co-author of the famous Oparin-Haldane
theory of biogenesis, which emphasized law-like aspects of the
process. This was in complete accordance with his philosophical
and methodological principles, and enabled him to lay down the
foundations of what is today often called future studies as well
(Clark 1968; Adams 2000).

\begin{figure}
\psfig{file=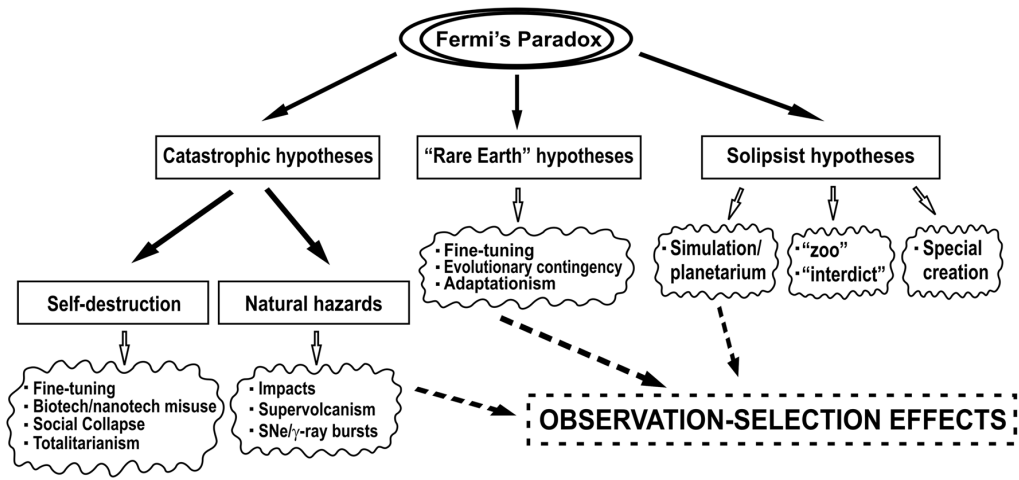,width=14cm} \caption{The proposed
high-level classification of the solutions to FP. In an extremely
simplified form, the respective replies to Fermi's question
\textit{Where is everybody?\/} by proponents of solipsist, "Rare
Earth" and (neo)catastrophic hypotheses are "They are here",
"They do not exist", and "They have been prevented from
coming/manifesting yet". Only a small subset of proposed
hypotheses is shown as examples in each category. \label{slidva}}
\end{figure}

An important novelty here in comparison to previous SETI reviews
is the necessity of taking into account hitherto unrecognized
possibilities, especially the Haldanian notion of {\bf
postbiological} evolution, prompted by Moore's Law and the great
strides made in the cognitive sciences. For instance, the great
historian of science Steven J. Dick (2003) cogently writes:

\begin{quote}
But if there is a flaw in the logic of the Fermi paradox and
extraterrestrials {\bf are} a natural outcome of cosmic
evolution, then cultural evolution may have resulted in a
postbiological universe in which machines are the predominant
intelligence. This is more than mere conjecture; it is a
recognition of the fact that cultural evolution - the final
frontier of the Drake Equation - needs to be taken into account
no less than the astronomical and biological components of cosmic
evolution. [emphasis in the original]
\end{quote}

It is easy to understand the necessity of redefining SETI studies
in general and our view of Fermi's Paradox in particular in this
context. For example, postbiological evolution makes those
behavioral and social traits like territoriality or expansion
drive (to fill the available ecological niche) which are---more
or less successfully---"derived from nature" lose their
relevance. Other important guidelines must be derived which will
encompass the vast realm of possibilities stemming from the
concept of postbiological evolution.

\section{Solipsist Solutions}

The subtitle refers to a classic 1983 paper of Sagan and Newman
criticizing Tipler's (1980, 1981) skepticism toward SETI studies
based on Fermi's Paradox (FP) and strengthened by the idea of
colonization via von Neumann probes. Here, however, we would like
to investigate solipsist solutions to FP in a different -- and
yet closer to the usual -- meaning.

Solipsist solutions reject the premise of FP, namely that there
are no extraterrestrial civilizations either on Earth or
detectable through our observations in the Solar System and the
Milky Way thus far. On the contrary, they usually suggest that
extraterrestrials are or have been present in our vicinity, but
that the reasons for their apparent absence lie more with our
observations and their limitations than with the real
state-of-affairs.

Of course, this has long been the province of the lunatic fringe
of science (either in older forms of occultism or more modern
guise of "ufology"), but to neglect some of these ideas for that
reason is to give the quacks too much power. Instead, we need to
consider all the alternatives, and these clearly form
well-defined, albeit often provably wrong or undeveloped ideas.
Hypotheses in this class serve another important role: they remind
us of the magnitude of the challenge posed by FP to our naive
worldview -- and they should be evaluated in this light. Some of
the solipsist hypotheses discussed at least half-seriously in the
literature are the following (listed in rough order from less to
more viable ones):

\begin{itemize}
\item
Those who believe {\bf UFOs} are of extraterrestrial intelligent
origin quite clearly do not have any problem with FP (e.g., Hynek
1972; for a succinct historical review see Chapter 6 of Dick
1996). The weight of evidence obviously tells otherwise.

\item
As far as it can be formulated as a hypothesis, traditional views
of {\bf special creation} of Earth and humanity belong to this
class. The most valiant attempt in this direction has been made,
as already mentioned, by Alfred Russell Wallace (1903), who
argued for the key role of "cosmic mind" in the grand scheme of
things and on the basis of a teleological (mis)interpretation of
the then-fashionable model of the universe similar to the
classical Kapteyn universe. As discussed in detail by Crow
(1999), such views were occasionally dressed in the garb of
traditional theology (especially of Christian provenance), but
the association is neither logically nor historically necessary
(see also Dick 2000, 2003). Today, this way of looking at the
problem of life and intelligence beyond Earth is abandoned in
most mainstream theologies (William Lane Craig, personal
communication).\footnote{Special creation, however, possesses
some methodological similarities to the "rare Earth" hypotheses
as well; see Section 6 below.}

\item
The {\bf Zoo hypothesis} of Ball (1973) and the related {\bf
Interdict hypothesis} of Fogg (1987) suggest that there is a
uniform cultural policy of advanced extraterrestrial civilization
to avoid any form of contact (including having visible
manifestations) with the newcomers to the "Galactic Club". The
reasons behind such a behavior may be those of ethics, prudence
or practicality (Deardorff 1987). In each case, these do not
really offer testable predictions (if the extraterrestrial
civilizations are sufficiently powerful, as suggested by the age
difference in \ref{tri}), for which they have been criticized by
Sagan, Webb and others. As a consequence, a {\bf "leaky"
interdict} scenario is occasionally invoked to connect with the
alleged extraterrestrial origin of UFOs (Deardorff 1986), which
is clearly problematic.

\item
The {\bf Directed panspermia} scenario/hypothesis of Crick and
Orgel (1973) suggests that Earth was indeed visited in the
distant past with very obvious consequences -- namely, the
existence of life on Earth! Those two famous biochemists proposed
-- partly tongue-in-cheek, but partly to point out the real
problems with the then theories of biogenesis -- that our planet
has been intentionally seeded with microorganisms originating
elsewhere. In other words, we are aliens ourselves! This motive
has been extensively used in fiction (e.g., Lovecraft 2005
[1931]). It is very hard to see how we could ever hope to test
the hypothesis of directed panspermia, in particular its
\textit{intentional\/} element.

\item
The {\bf Planetarium hypothesis} of Baxter (2000) suggests that
our astronomical observations do not represent reality, but a
form of illusion, created by an advanced technological
civilization capable of manipulating matter and energy on
interstellar or Galactic scales. For a fictional description of
this scenario, see Reynolds (2004).

\item
The {\bf Simulation hypothesis} of Bostrom (2003), although
motivated by entirely different reasons and formulated in a way
which seemingly has nothing to do with FP, offers a framework in
which FP can be naturally explained. Bostrom offers a Bayesian
argument for why we might rationally think we live in a computer
simulation of an advanced technological civilization inhabiting
the "real" universe. This kind of argument has a long
philosophical tradition, going back at least to Descartes'
celebrated second \textit{Meditation\/} discussing the level of
confidence we should have about our empirical knowledge (for an
interesting recent review, see Smart 2004). Novel points in
Bostrom's presentation include invoking Moore's Law in order to
suggest that we might be technologically closer to the required
level of computing sophistication than we usually think, as well
as adding a Bayesian conditioning on the number (or sufficiently
generalized "cost" in resources) of such "ancestor-simulations"
as he dubs them. It is trivial to see how FP is answered under
this hypothesis: extraterrestrial civilizations are likely to be
simply beyond the scope of the simulation in the same manner as,
for example, present-day simulations of the internal structure of
the Sun neglect the existence of other stars in the universe.
\end{itemize}
\vspace{0.3cm}

It is difficult to objectively assess the value of solipsist
hypotheses as solutions to FP. Most of them are either untestable
in principle like the eponymous metaphysical doctrine, or testable
only at the limit of very long temporal and spatial scales, so
that they do not belong to the realm of science, conventionally
understood. In other words, they violate a sort of "naive"
realism which underlies practically the entire scientific
endeavor. Their proponents are likely to retort that the issue is
sufficiently distinct from other scientific problems to justify a
greater divergence of epistemological attitudes -- but this is
rather hard to justify when one could still pay a smaller price.
For instance, one could choose to abandon Copernicanism, like the
Rare Earth theorists (Section 6), or one might abandon gradualism
(which has been discredited in geo- and planetary sciences
anyway) and end up with a sort of neocatastrophic hypothesis
(Section 7).

Some of them, but not all, violate the non-exclusivity
requirement as well; this is, for instance, obvious in the Zoo,
Interdict or Planetarium scenarios, since they presume a
large-scale cultural uniformity. This is not the case, however,
with the Simulation hypothesis, since the simulated reality is
likely to be clearly designed and spatially and temporally
limited. Directed panspermia has some additional problems --
notably the absence of any further manifestations of our "parent
civilization", in spite of its immense age. If they became
extinct in the meantime, what happened with the other seeded
planets? Copernican reasoning suggests that we should expect
evolution to occur faster at some places than on Earth (and, of
course, slower at other sites as well) -- where, then, are our
interstellar siblings?

Observation selection effects are important ingredient in at
least some of these hypotheses. The Directed panspermia scenario
could, for instance, be linked with a curious puzzle posed
recently by Olum (2004), which also helps to illustrate the
intriguing interplay between modern cosmology and astrobiology.
Starting from the assumption of an infinite universe (following
from the inflationary paradigm), Olum conjectures that there are
civilizations much larger than ours (which currently consists of
about $10^{10}$ observers). The spatial extent and amount of
resources at the disposal of such large civilizations would lead,
in principle, to much larger number of observers (for example,
$10^{19}$ observers in a Kardashev Type III civilization).  Now,
even if 99\% of all existing civilizations are small ones similar
to our own, anthropic reasoning suggests that the overwhelming
probabilistic prediction is that we live in a large civilization.
This prediction is spectacularly unsuccessful on empirical
grounds; with a probability of such failure being about $10^{-8}$,
something is clearly wrong here. Olum offers a dozen or so
hypothetical solutions to this alleged conflict of anthropic
reasoning with cosmology, one of them being the possibility that
we are indeed part of a large civilization without being aware of
that fact. The Directed panspermia hypothesis can be regarded as
operationalization of that option. There are several systematic
deficiencies in Olum's conclusions (Ho and Monton 2005,
\'Cirkovi\'c 2006), but in any case the very fact that some form
of the principle of indifference and the counting of observers is
used in this discussion shows how closely the theory of
observation selection effects (cf.\ Bostrom 2002) is tied up with
issues at the very heart of FP.

We mention the solipsist hypotheses mostly for the sake of
logical completeness, since they are in any case a council of
despair. If and when all other avenues of research are exhausted,
we could always turn toward these hypotheses. Still, this neither
means that they are all of equal value nor it should mislead us
into thinking that they are necessarily improbable for the reason
of desperation alone. Bostrom's simulation hypothesis might,
indeed, be quite probable, given some additional assumptions
related to the increase in our computing power and decrease of
information-processing cost. The Directed panspermia hypothesis
could, in principle, get a strong boost if, for instance, the
efforts of NASA and other human agencies aimed at preventing
planetary contamination (e.g., Rummel 2001, Grinspoon 2003), turn
out to be unsuccessful, thus unintentionally setting off
biological evolution on other Solar System bodies. Finally,
solipsist hypotheses need not worry about evolutionary
contingency or generic probabilities of biogenesis or noogenesis,
unlike the other contenders.

Jumping ahead, a clearly non-exclusive solution to FP obeying all
methodological desiderata has not, in general, been found thus
far. Even the most objective, mathematical studies, such as the
one of Newman and Sagan, were compelled to, somewhat resignedly,
conclude that "[i]t is curious that the solution to the problem
'Where are they?' depends powerfully on the politics and ethics
of advanced societies" (Newman and Sagan 1981, p. 320). There is
something deeply unsatisfactory about this sort of answer. It is
especially disappointing to encounter it after a lot of
mathematical analysis by the same authors, and keeping in mind by
now more than half a century of sustained and often carefully
planned and executed SETI efforts. This circumstance, as well as
occasional (sub)cultural and even political appeal, explains why
solipsist hypotheses are likely to reappear from time to time in
the future.

\section{"Rare Earth" Solutions} \label{rare}

This class of hypotheses is based upon the celebrated book
\textit{Rare Earth\/} by Peter Ward and Donald Brownlee, whose
appearance in 2000 heralded the birth of the new astrobiological
paradigm. They have expounded  a view that while simple microbial
life is probably ubiquitous throughout the Galaxy, complex
biospheres, like the terrestrial one, are very rare due to the
exceptional combination of many distinct requirements. These
ingredients of the {\bf Rare Earth hypothesis} (henceforth REH)
are well-known to even a casual student of astrobiology:

\vspace{0.3cm}

\begin{itemize}
\item
Circumstellar habitable zone: a habitable planet needs to be in
the very narrow interval of distances from the parent star.

\item
"Rare Moon": having a large moon to stabilize the planetary axis
is crucial for the long-term climate stability.

\item
"Rare Jupiter": having a giant planet ("Jupiter") at the right
distance to deflect much of the incoming cometary and asteroidal
material enables sufficiently low level of impact catastrophes.

\item
"Rare elements": Radioactive $r$-elements (especially U and Th)
need to be present in the planetary interior in sufficient amount
to enable plate tectonics and functioning of the carbon-silicate
cycle.

\item
"Rare Cambrian-explosion analogs": the evolution of complex
metazoans requires exceptional physical, chemical and geological
conditions for episodes of sudden diversification and expansion
of life.
\end{itemize}
\vspace{0.3cm}

Each of these requirements is \textit{prima facie\/} unlikely, so
that their combination is bound to be incredibly rare and probably
unique in the Milky Way. In addition, Ward and Brownlee break new
ground by pointing out the importance of hitherto downplayed
factors, like the importance of plate tectonics, inertial
interchange events, or "Snowball Earth" episodes of global
glaciation for the development of complex life. In many ways, REH
has become somewhat of a default position in many astrobiological
circles, and -- since it predicts the absence of rationale for
SETI -- a mainstay of SETI scepticism. Thus, its challenge to
Copernicanism has been largely accepted (although, as argued
below, there are lower prices to be paid on the market of ideas)
as sound in mainstream astrobiology. Particular Rare Earth
hypotheses (insofar as we may treat them as separate) are
difficult to assess lacking first-hand knowledge of other
Earthlike planets, but some of the difficulties have been exposed
in the literature thus far.

For instance, the famous argument about Jupiter being the optimal
"shield" of Earth from cometary bombardment has been brought into
question by recent work of Horner and Jones (2008, 2009) who use
numerical simulation to show that the off-handed conclusion that
Jupiter acts as a shield against bombardment of inner Solar
System planets is unsupported. Moreover, they conclude "that such
planets often actually increase the impact flux greatly over that
which would be expected were a giant planet not present." If
results of Horner and Jones withstand the test of time and further
research, it is hard to imagine a more detrimental result for the
entire Rare Earth paradigm.

This example highlights the major problem with REH. In supposing
how the state-of-affairs could be different, Rare Earth theorists
assume simple, linear change, not taking into account the
self-organizing nature of the relevant physical systems. The
example of Jupiter is again instructive, since \textbf{asking
about the fate of Earth in the absence of Jupiter is
self-contradictory}: Earth is a part of the complex system which
includes Jupiter as a major component, so there are no guarantees
that Earth would have existed at all if Jupiter were not present.
Even if it had existed, we would have to account for many other
differences between that particular counterfactual situation and
the actual one, so the question to what degree it is justified to
call such a body "Earth" would be very pertinent.

Another important methodological problem for the "rare Earth"
hypotheses is that at least in some respects they are equivalent
to the doctrines openly violating naturalism, e.g., creationism.
This similarity in style rather than in substance has been most
forcefully elaborated by Fry (1995), as mentioned above. If one
concludes that the probability of biogenesis -- even under
favorable physical and chemical preconditions -- is astronomically
small, say $10^{-100}$, but one still professes that it was
completely natural event,\footnote{Even smaller probabilities
have been occasionally cited in the literature. Thus, Eigen
(1992) cites the probability of random assembly of a polymer with
a thousand nucleotides corresponding to a single gene as 1 part
in $10^{602}$. This sort of "superastronomical" number has led
Hoyle and Wickramasinghe (1981, 1999) to invoke either an eternal
universe -- in contradiction with cosmology -- or a creative
agency. The (in)famous metaphor of the random assembly of a
"Boeing 747" out of a junkyard, cited by Sir Fred Hoyle, nicely
expresses this sort of desperation, which has, luckily enough,
been overcome in the modern theories of biogenesis.} than a
curious situation arises in which an opponent can argue that
supernatural origin of life is clearly more plausible hypothesis!
Namely, even a fervent atheist and naturalist could not
rationally claim that her probability of being wrong on this
metaphysical issue is indeed smaller than $10^{-100}$, knowing
what we know on the general fallibility of human cognition.
According to the dominant rules of inference, we would have been
forced to accept the creationist position, if no other hypothesis
were present (Hoyle and Wickramasinghe 1999)! Now, REH in strict
sense avoids this problem by postulating ubiquitous simple life
(actually implying a high probability of biogenesis
\textit{ceteris paribus\/}). However, {\bf if} the continuity
thesis applies further along "Haldane's ladder" -- specifically,
to origin of complex metazoans and to noogenesis -- an analogous
argument is perfectly applicable to REH. If the probability of
evolutionary pathways entering a small region of biological
morphospace describing intelligent tool-making species is
astronomically small, and one still maintains that it has occurred
on Earth naturalistically, as a "happy accident", one is open to
the same criticism as Fry brought forward in case of biogenesis.
Obviously, this necessitates further research in evolutionary
biology, cognitive sciences and philosophy.

There are other hypotheses for resolving FP which violate
Copernicanism. The idea of Wesson (1990) that it is cosmology
which limits the contact between civilizations in the universe
also belongs to this category. It implies that the density of
civilizations is so low that only a few are located within our
cosmological horizon. However, this is just begging the question,
since such an extreme low density of inhabited sites -- less than
1 Gpc$^{-1}$, say -- is not only un-Copernican, but clearly
requires some additional explanatory mechanism. It may consist in
biological contingency or the rarity of the Cambrian-explosion
analogs, or any number of other instances invoked by the
proponents of REH, but it is clearly necessary.

On the other hand, no further explanation is necessary for the
{\bf adaptationist version} of REH, which in this case could truly
be dubbed the {\bf "rare mind" hypothesis}. It has been hinted at
by Raup (1992), but developed in more detail in  the novel {\it
Permanence\/} by the Canadian author Karl Schroeder (2002). A
detailed discussion of this particular solution to FP is given in
\'Cirkovi\'c (2005). This intriguing hypothesis uses the
prevailing adaptationist mode of explanation in evolutionary
biology to argue that conscious tool-making and
civilization-building are ephemeral adaptive traits, like any
other in the living world. Adaptive traits are bound to disappear
once the environment changes sufficiently for any selective
advantage which existed previously to disappear. In the long run,
intelligence is bound to disappear, as its selective advantage is
temporally limited by ever-changing physical and ecological
conditions. The outcome of cultural evolution at limits of very
long timescales is a reversion to direct, non-technological
adaptation -- similar to the suggestion of Raup that animals on
other planets may have evolved, by natural selection, the ability
to communicate by radio waves (and, by analogy, at least some of
the other traits we usually think about as possible only within
the conscious civilization). This form of downgrading of the role
of consciousness -- present in many circles of contemporary
philosophy of mind and cognitive science -- is beautifully
illustrated in the controversial book of Julian Jaynes
(1990).\footnote{A particularly thought-provoking section (pp.\
36-41) of the first chapter of Jaynes' disturbing book is
entitled "Consciousness Not Necessary for Thinking". See also N\o
rretranders 1999.}

There are many difficulties with the adaptationist hypothesis.
For instance, its insistence on adaptationism at all times is a
form of inductivist fallacy. As in earlier times inductivists
argued that it is natural to assume a meta-rule of inference
along the lines of "the future will resemble the past", thus there
is a creeping prejudice that the present and future modes of
evolution need to be the same as those leading to the present
epoch. This is a consequence of the present-day idolatry of
adaptation: the almost reflex assumption that any evolution has
to be adaptationist (e.g., Dennett 1995; for a criticism, see
Ahouse 1998). In spite of such fashionable views like
evolutionary psychology/behavioral ecology/sociobiology, there is
no reason to believe that all complex living systems evolve
according to the rules of functionalist natural selection, and
not, for instance, in a Lamarckian, orthogenetic or saltationist
manner. Besides, even if {\bf all} Gyr-old civilizations are now
extinct, what about their astroengineering traces and
manifestations? For a detailed review of further problematic
issues with this intriguing hypothesis, see \'Cirkovi\'c,
Dragi\'cevi\'c and Beri\'c-Bjedov (2005).

\section{(Neo)Catastrophic Solutions} \label{catas}

This is the most heterogeneous group, containing both some of the
oldest and newest speculations on the topic. Before we review
some of the main contenders, it is important to emphasize that
the prefix "neo" is used almost reflexively with this mode of
thinking for historical reasons. The defeat of the "classical",
19th-century catastrophism of figures such as Cuvier, Orbigny, de
Beaumont, Agassiz or Sedgwick in the grand battle with the
gradualism of Charles Lyell and his pupils (including Charles
Darwin) imposed a lasting stigma on views which were perceived as
beloging to this tradition of thought. This has clearly impeded
the development of geosciences (see historical reviews in Raup
1991, Huggett 1997, Palmer 2003). In addition, the association of
catastrophism with the pseudo-scientific (although often
thought-provoking!) views of Immanuel Velikovsky in 1950s and
1960s has brought an additional layer of suspicion upon the label
itself (for a review of the Velikovskian controversy, see Bauer
1984). Thus, the resurgence of catastrophism after 1980 and the
discovery by Alvarez and collaborators that an
asteroidal/cometary impact was the physical cause of the
extinction of ammonites, dinosaurs and other species at the
Cretaceous/Tertiary boundary 65 Myr ago (Alvarez et al. 1980) is
often referred to as 'neocatastrophism'.

\vspace{0.3cm}

\begin{itemize}
\item
Classical {\bf nuclear self-destruction hypothesis} was, perhaps
more prevalent during the Cold War era (cf.\ von Hoerner 1978),
but ephemeral cultural changes in our recent history should not
really modify the prior probability for this dramatic possibility.
Problems with the exclusive nature of such a hypothesis --
considering the fact that social and political developments on
habitable planets throughout the Galaxy are quite unlikely to be
correlated -- are obvious.

\item
{\bf Self-destruction options} have multiplied in the meantime,
since the spectrum of potentially destructive technologies in
human history have recently broadened. This now includes misuse of
biotechnology (including bioterrorism), and is likely to soon
include misuse of nanotechnology, artificial intelligence, or
geoengineering (see reviews in Bostrom and \'Cirkovi\'c 2008;
\'Cirkovi\'c and Cathcart 2004). If most technological societies
in the Galaxy self-destruct through any of these -- or other
conceivable -- means, this would be an explanation for the "Great
Silence". Quite clearly, the same qualms about exclusivity apply
as above.

\item
{\bf Ecological holocaust}: Our Solar System and surrounding
parts of GHZ belong to a "postcolonization wasteland", a bubble
created by rapid expansion and exhaustion of local resources on
the part of early advanced technological civilizations (Stull
1979; Finney and Jones 1985). Since colonization front is likely
to be spherically symmetric (or axially symmetric when the
vertical boundaries of the Galactic disk are reached), they will
tend to leave vast inner area exhausted. If the parameters
describing the rates of expansion and natural renewal of
resources are in a particular range of values, it is possible
that younger civilizations will find themselves in a  This
hypothesis has been recently revived in numerical models of
Hanson (1998b), showing that in some cases fairly plausible
initial conditions will lead to "burning of the cosmic commons",
i.e. catastrophic depletion of usable resources in a large volume
of space. This is rather controversial {\bf as a solution to FP}
since, apart from some fine-tuning, it still does not answer the
essential question: where did the "precursors" go and why we do
not perceive their immensely old astro-engineering signatures?
They have either become extinct (thus begging the question and
requiring another layer of explanation) or changed into something
else (see the {\bf Transcedence} item below). However, this
hypothesis is non-exclusive (since the volume of space within the
ancient colonization front is large) and it does make some
well-defined predictions as far as renewal of resources and the
traces of possible previous cycle of their depletion in the Solar
vicinity are concerned.

\item
{\bf Natural hazards}: The risk of cometary/asteroidal bombardment
(e.g., Clube and Napier 1984, 1990, Chyba 1997), supervolcanism
(Rampino 2002), nearby supernovae (Terry and Tucker 1968, Gehrels
et al.\ 2003) or some other, more exotic catastrophic process
(Clarke 1981) might be in general much higher than we infer from
the recent history of Earth. These natural hazards are much
likelier to break the evolutionary chain leading to the emergence
of intelligent observers, so we should not wonder why we do not
perceive manifestations of older Galactic communities. For
instance, one well-studied case is the system of the famous
nearby Sun-like star Tau Ceti, which contains both planets and a
massive debris disk, analogous to the Solar System's Kuiper belt.
Modeling of Tau Ceti's dust-disk observations indicate, however,
that the mass of the colliding bodies up to 10 kilometers in size
may total around 1.2 $M_\oplus$, compared with 0.1 $M_\oplus$
Earth-masses estimated to be in the Solar System's
Edgeworth-Kuiper Belt (Greaves et al. 2004). It is only
reasonable to conjecture that any hypothetical terrestrial planet
of this extrasolar planetary system is subjected to much more
severe impact stress than Earth has been during the course of its
geological and biological history.\footnote{For a good recent
introduction to the complex topic of the relationship between
catastrophes and habitability, see Hanslmeier (2009).}

\item
The {\bf Phase-transition hypotheses} (Annis 1999a, \'Cirkovi\'c
2004b, \'Cirkovi\'c and Vukoti\'c 2008) offers a plausible
astrophysical scenario for a delay in the emergence of
intelligent observers and their technological civilizations based
on the notion of a global regulation mechanism. Such a mechanism
could occasionally reset astrobiological "clocks" all over GHZ
and in a sense re-synchronize them. This is is a prototype {\bf
disequilibrium} astrobiological hypothesis: there is no Fermi's
paradox, since the relevant timescale is the time elapsed since
the last "reset" of astrobiological clocks and this can be
substantially smaller than the age of the Milky Way or the age
difference in (3). Annis suggests that gamma-ray bursts
(henceforth GRBs), whose cosmological and extremely energetic
nature is now increasingly understood (e.g., M\'{e}sz\'{a}ros
2002, Woosley and Bloom 2006) serve as such catastrophic reset
events when they occur in our home Galaxy. The astrobiological
significance of GRBs has recently been the subject of much
research (Thorsett 1995, Scalo and Wheeler 2002, Thomas et al.
2005, 2008, Galante and Horvath 2007). The discussion of other
conceivable regulation mechanisms is given by Vukoti\'c and
\'Cirkovi\'c (2007, 2008). In general, this hypothesis leads to
the situation schematically envisioned in Fig. 3: where we are
within the temporal window of a "phase transition" -- from an
essentially dead place, the Galaxy will be filled with
intelligent life on a timescale similar to $t_{FH}$.

\item
The {\bf Deadly Probes hypothesis:} A particularly disturbing
version of Tipler's (1980, 1981) \textit{reductio ad absurdum\/}
scenario presumes that self-replicating von Neumann probes are not
peaceful explorers or economically-minded colonizers, but
intentionally or accidentally created destructive weapons. This
might occur either due to malevolent creators (which in that case
would have to be the first or one of the first technological
civilizations in the Galaxy, close to the Lineweaver limit) or
through a random dysfunction ("mutation") in a particular
self-replicating probe which has passed to its "offspring". In
both cases, it seems that the originators of the probes have
vanished or are in hiding, while the Galaxy is a completely
different (and more hostile) ecological system than is usually
assumed. Depending on the unknown mode of operation of
destructive von Neumann probes, they might be homing in on the
sources of coherent radio emission (indicating a young
civilization to be eliminated) or might be automatically sweeping
the Galaxy in search for such adversaries. Brin (1983) concludes
that this is one of only two hypotheses which maintain wholesale
agreement with both observation and non-exclusivity. In the realm
of fiction, this hypothesis has been topic of novels by Fred
Saberhagen (1998), Gregory Benford (1977, 1983) and Alastair
Reynolds (2002).

\item
{\bf "Freedom is slavery"}: If all civilizations, instead of
self-destructing, slip into permanent totalitarianism (perhaps in
order to avoid self-destruction or other global catastrophic
risks; see Caplan 2008), this could also dramatically decrease the
contact cross-section. This Orwellian State is quite
disinterested in the external universe; even if it were willing
to communicate, its paranoid nature would have made any
opportunity for contact orders of magnitude more difficult. For a
gruesomely dramatic description of this possibility see
\textit{Fiasco\/} (Lem 1987). On the other hand, it is
conceivable that at least some totalitarian states would actually
engage in aggressive interstellar expansion, even if through
releasing the deadly probes sketched above. Here, as elsewhere,
we might have a case for synergy of different FP solutions.

\item
The {\bf Transcendence hypothesis}: Advanced technological
civilizations have neither destroyed themselves nor spread through
the Galaxy, but have transformed themselves into "something else",
not recognizable as a civilization and certainly not viable as a
SETI target. Historically, this has been the first solution to FP,
offered by Konstantin Tsiolkovsky who posed the paradox in the
first place. Tsiolkovsky, under the influence of his teacher N.
F. Fedorov and other Russian cosmists,  concluded that the only
reason why we do not perceive manifestations of much older
civilizations is their evolving into a form of "superreason" with
near-godly powers and, presumably, inconceivable interests
(Tsiolskovsky 1933; see also Lytkin et al.\ 1995, Lipunov 1997).
The ideas of Tsiolkovsky have some similarities with the Zoo
hypothesis of Ball (1973), discussed above. Today, it is often
formulated in terms of a "technological Singularity", the concept
envisioned by Stanislaw Ulam and I. J. Good, and popularized in
the 1990s by mathematician and author Vernon Vinge (e.g., Vinge
1986, 1991, 1993; Kurzweil 2005). Smart's (2007) concept of the
"Universal Transcension" is a variation of this idea. There are
two reasons why is this vague family of scenarios classified
here, with other catastrophic hypotheses. First, the external
appearance of a transcending event or process might be
catastrophic; fictional precursor of the hypothesis, Vinge (1986)
hints at such scenario. More importantly, one of the few common
claims for the transcedence scenarios is that of sudden, discrete
change, obviously antithetical to gradualism.
\end{itemize}

\vspace{0.3cm}

As the Cold War cultural pessimism retreated, neocatastrophic
hypotheses obtained a strong boost from the resurgence of
catastrophism in Earth and planetary science, as well as in
astrobiology. Following the seminal work of Alvarez et al.
(1980), we have become aware that global catastrophes played a
very significant role in the evolution of the terrestrial
biosphere (e.g., Jablonski 1986, Raup 1991, Courtillot 1999, Erwin
2006). Moreover, some of the actual catastrophes whose traces are
seen in the terrestrial record are of astrophysical origin,
emphasizing the new paradigm according to which the Solar System
is an open system, strongly interacting with its Galactic
environment (e.g., Clube and Napier 1990, Leitch and Vasisht 1998,
Shaviv 2002, Melott et al. 2004, Pavlov et al. 2005, Gies and
Helsel 2005). This neocatastrophist tendency is present in modern
research on biogenesis (e.g., Raup and Valentine 1983, Maher and
Stevenson 1988), and even in the debates on the evolution of
humanity (Rampino and Self 1992, Ambrose 1998, Bostrom and
\'Cirkovi\'c 2008), but all of its ramifications have not yet
been elucidated in any detail. The major feature of these
solutions is the abandonment of the classical gradualist dogma
that "the present is key to the past" and the acknowledgement
that sudden, punctuated changes present a major ingredient in
shaping both the Earth's and the Milky Way's astrobiological
history (or "landscape"; cf. Vukoti\'c and \'Cirkovi\'c 2008).

Intuitively, it seems clear that any form of catastrophic event
affecting planetary biospheres in the Milky Way will reduce the
hypothetical extraterrestrial civilizations' ages and thus reduce
the tension inherent in FP. If such events are spatially and
temporally uncorrelated -- as in the "mandatory" nuclear
self-destruction hypothesis or the totalitarian scenario -- such
an explanation is obviously low on the non-exclusivity scale. In
contrast, hypotheses with correlated events -- such as "deadly
probes" or phase-transitions -- fare much better here. In some
cases it is still impossible to estimate how tightly correlated
some of the postulated events might be; this applies in
particular to the transcendence-type scenarios, where the extent
and nature of the "Singularity" remains a
mystery.\footnote{Consequently, it is impossible to state
confidently whether the transcendence hypotheses resolve FP,
i.e., what additional assumptions are necessary for this rather
vague concept to be a viable solution. On the other hand, the
obvious -- and rather dramatic -- importance of this scenario for
future studies remains a strong motivation for further research.}

Among the non-exclusive hypotheses, the phase-transition model
harbours an advantage in comparison to the "deadly probes"
scenario, since we understand possible dynamics of the global
regulation mechanisms. Moreover, global catastrophic events
affecting large parts of GHZ will tend to reset many local
astrobiological clocks nearly simultaneously, thus significantly
decreasing the probability of the existence of extremely old
civilizations, in accordance with Annis' scenario. In both of
these hypotheses, however, it is possible that pockets of old (in
effective, astrobiological terms) habitable sites would remain,
either through the purely stochastic nature of lethal regulation
mechanisms, or through the dysfunctional mode of operation of
destructive von Neumann probes.

Predictions of these two hypotheses and their ramifications for
the ongoing SETI projects could not differ more dramatically.
While the "deadly probes" scenario is particularly bleak and
offers no significant prospect for SETI, punctuation of the
astrobiological evolution of the Milky Way with large-scale
catastrophes affecting a significant fraction of GHZ would,
somewhat counterintuitively, have the net effect of strengthening
the rationale for our present-day SETI efforts. Namely, as the
secular evolution of the regulation mechanisms leads to the
increase in the average astrobiological complexity (Fig. 3), we
might expect that more and more civilizations enter the "contact
window" and join efforts in expansion towards Kardashev's Type
III status.

\begin{figure}
\psfig{file=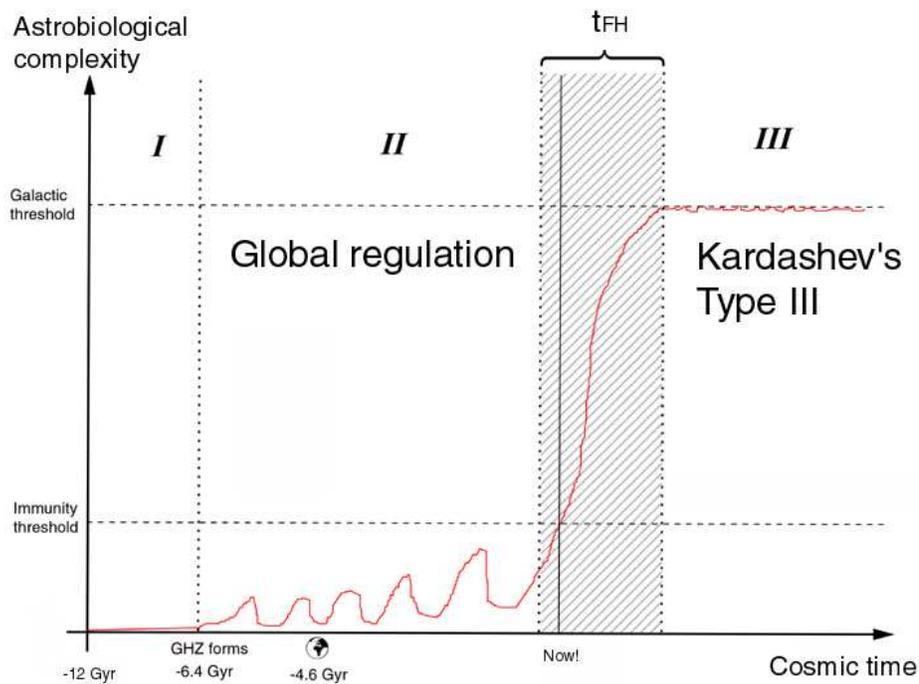,width=13cm} \caption{Very simplified
scheme of the phase-transition hypotheses (from \'Cirkovi\'c and
Vukoti\'c 2008): an appropriately defined astrobiological
complexity will tend to increase with time, but the increase will
not become monotonous until a particular epoch is reached.
\label{pic2}}
\end{figure}

\section{Other Solutions}

A small number of hypotheses have been proposed which do not fall
easily into any of the broad categories described above. Although
the total variation of approaches to FP is already stupendous, it
is remarkable how a small number of remaining ideas defy
inclusion within the general philosophical categories we have so
far discussed.

For instance, Landis (1998) and Kinouchi (2001) have investigated
the dynamics of interstellar colonization which, under some
particular assumptions, can leave large bubbles of empty space
surrounded by colonized regions. This phenomenon appears in the
context of condensed-matter physics as \textit{persistence}. An
obvious weakness of this hypothesis is that it still implies
cultural uniformity regarding the dynamical parameters of
colonization, which violates the non-exclusivity requirement. In
addition, we would expect to detect either extraterrestrial
signals coming from outside of the local non-colonized bubble, or
to detect manifestations of Gyr-older technological societies
even in the absence of the direct presence of extraterrestrials
in the Solar System or in its vicinity.

A similar approach has been favored in the numerical simulations
of Bjork (2007), although the timescales obtained in his model
are quite short in comparison with (\ref{tri}), even with his
explicit rejection of self-reproducing probes, thus being more in
line with the older calculations of Hart (1975), Jones (1976,
1981) and Newman and Sagan (1981). Bjork concludes, rather too
optimistically, that FP could be resolved by the statement that
"[w]e have not yet been contacted by any extraterrestrial
civilizations simple because they have not yet had the time to
find us." In view of timescale (\ref{tri}) it is clearly wrong as
long as we do not postulate some additional reason for the delay
in the initiation Galactic exploration.

The approach of \'Cirkovi\'c and Bradbury (2006; see also
\'Cirkovi\'c 2008) offers an alternative solution based on the
assumption that most or all advanced technological societies will
tend to optimize their resource utilization to an extreme degree.
It could be shown that such optimization will ultimately be
limited by the temperature of interstellar space -- and that
temperature decreases with increased galactocentric distance in
the Milky Way (towards the ideal case of the CMB temperature of
about 2.7 K, achievable only in intergalactic space). The logical
conclusion is that most of the advanced technological species
(which will be most likely postbiological, consisting of
intelligent machines or uploaded minds; cf. Dick 2003) will
migrate towards the outer rim of the Galaxy, far from the
star-formation regions, supernovae and other energetic
astrophysical events, in order to process information most
efficiently. This solution modestly violates the non-exclusivity
requirement, depending on how universally valid is the assumption
of resource-optimization as the major motivator of advanced
extraterrestrial societies.

Not surprisingly, some of these ideas have been prefigured in a
loose form within the discourse of science fiction. Karl
Schroeder in \textit{Permanence\/} not only formulated the
above-mentioned adaptationist answer to Fermi's question, but
also envisaged the entire Galaxy-wide ecosystem based on brown
dwarfs (and the halo population in general) and a low-temperature
environment (Schroeder 2002). Most strikingly, the idea of an
advanced technological civilization inhabiting the outer fringes
of the Milky Way has been suggested -- though without the
thermodynamical rationale -- by Vernon Vinge in \textit{A Fire
upon the Deep\/} (Vinge 1991). Vinge vividly envisages "Zone
boundaries" separating dead and low-tech environments from the
truly advanced societies inhabiting regions at the boundary of
the disk and high above the Galactic plane. This is roughly
analogous to the low- temperature regions \'Cirkovi\'c and
Bradbury (2006) outlined as the most probable Galactic
technological zone.

It has been claimed in the classical SETI literature that the
interstellar migrations will be forced by the natural course of
stellar evolution (Zuckerman 1985). However, even this
"attenuated" expansionism -- delayed by an order of $10^9$ years
-- is actually unnecessary, since naturally occurring
thermonuclear fusion in stars is extremely inefficient energy
source, converting less than 1\% of the total stellar mass into
potentially useable energy. A much deeper (by at least an order of
magnitude) reservoir of useful energy is contained in the
gravitational field of a stellar remnant (white dwarf, neutron
star or black hole), even without already envisaged stellar
engineering (Criswell 1985, Beech 2008). A highly optimized
civilization will be able to prolong utilization of its
astrophysically local resources to truly cosmological timescales.
The consequences for our conventional (that is, predominantly
empire-state) view of advanced societies have been encapsulated in
an interesting speculative paper by Beech (1990):

\begin{quote}
\noindent [A] star can only burn hydrogen for a finite time, and
it is probably safe to suppose that a civilisation capable of
engineering the condition of their parent star is also capable of
initiating a programme of interstellar exploration. Should they
embark on such a programme of exploration it is suggested that
they will do so, however, \textbf{by choice rather than by
necessitated practicality}. [emphasis M. M. \'C.]
\end{quote}

In brief, the often-quoted clich\'{e} that life fills all
available niches is clearly a \textit{non sequitur\/} in the
relevant context; thus, interstellar colonial expansion should
not be a default hypothesis, which it sadly is in most
SETI-related and far-future-related discourses thus far.

The {\bf sustainability solution} of Haqq-Misra and Baum (2009) is
related to the compact, highly-efficient model of advanced
extraterrestrial civilization postulated in Parkinson (2004),
\'Cirkovi\'c and Bradbury (2006), Smart (2007), and \'Cirkovi\'c
(2008). Haqq-Misra and Baum envision a situation in which
large-scale interstellar expansion is infeasible due to
sustainability costs (and perhaps dysgenic factors, similar to
the ones in Schroeder's adaptationist hypothesis), so that the
prevailing model would be a compact,
technologically-sophisticated "city-state" civilization, possibly
slowly expanding, but at rates negligible in comparison to the
expansion in either the Newman-Sagan-Bjork (no self-replicating
probes) or Tipler (with self-replicating probes) regimes.
Parkinson's (2004) {\bf containment scenario} offers a different
rationale for the predominance of the "city-states" over the
"interstellar empires", resulting in the same observed dearth of
interstellar empires. These hypotheses meet with the same
criticisms based on (i) the non-exclusivity and (ii) the lack of
astroengineering detection signatures considered above.

\section{Instead of Conclusions: A Puzzle for the 3. Millennium?}

The very fact that {\bf each wide class of answers to FP requires
abandoning one of the great methodological assumptions of modern
science} (solipsist solutions reject naive realism, "rare Earth"
solutions reject Copernicanism and neocatastrophic solutions --
gradualism) should give us pause.\footnote{We have assumed
naturalism throughout, in accordance with the proclaimed goal of
investigating to what degree FP remains {\bf un}resolved.} This
testifies to the toughness and inherent complexity of the puzzle.
In accordance with the strong position of REH in contemporary
astrobiology, our analysis shows that we should interpret it as a
challenge to Copernicanism. In the view of the present author, by
far the lowest price if paid through abandoning of gradualism,
which is anyway undermined by contemporary developments in the
geosciences, evolutionary biology and astronomy.

Gradualism, parenthetically, has not shone as a brilliant guiding
principle in astrophysics and cosmology either. It is well-known,
for instance, how the strictly gradualist (and from many points of
view methodologically superior) steady-state theory of the
universe of Bondi and Gold (1948), as well as Hoyle, has, since
the "great controversy" of the 1950s and early 1960s, succumbed
to the rival evolutionary models, now known as the standard ("Big
Bang") cosmology (Kragh 1996). Balashov (1994) has especially
stressed this aspect of the controversy by showing how deeply
justified was the introduction -- by the Big Bang cosmologists --
of events and epochs never seen or experienced. Similar arguments
are applicable in the nascent discipline of astrobiology, which
might be considered to be in an analogous state today as
cosmology was half a century ago (\'Cirkovi\'c 2004a).

This leads us to the practical issue of the ramifications of the
various hypotheses sketched above for practical SETI activities.
While solipsist hypotheses have nothing substantial to offer in
this regard, Rare Earth hypotheses obviate the very need for
practical SETI efforts. In the best case, we could expect to find
archaeological traces of some extinct Galactic civilization (as
per the adaptationist hypothesis). In contrast, most
neocatastrophic options offer support for SETI optimism, since
their proponents expect practically all extraterrestrial
societies to be roughly of the same effective age as
ours,\footnote{The qualification "effective" is required here
since in the case of arrested development (e.g., under the
totalitarianism scenario), the age of civilization is almost
irrelevant to its capacity for cosmic colonization.} and to be
our competitors for the Fermi-Hart-Tiplerian colonization of the
Milky Way. The price to be paid for bringing the arguments of
"optimists" and "pessimists" into accord is, obviously, the
assumption that we are living in a rather special epoch in
Galactic history -- i.e.\ the epoch of a phase transition. In any
case, it is clear that our choice of hypotheses for resolving FP
needs to impact our SETI efforts in a most direct way.

A related issue too complex to enter into here in more detail is
the inadequacy of most of the orthodox SETI projects thus far.
Radio listening for intentional messages, either intercepted or
specifically directed to young societies, has been a trademark of
orthodox SETI since the time of its "founding fathers" (Drake,
Morrison, Sagan, etc.) and it has demonstrated quite a strong
resilience to dramatic changes in other fields of learning over
the past four decades. Several issues touched upon in this review
strongly indicate that the conventional SETI (Tarter 2001, Duric
and Field 2003, and references therein), as exemplified by the
historical OZMA Project, as well its current counterparts such as
META, ARGUS, Phoenix, SERENDIP/Southern SERENDIP -- and notably
those conveyed by NASA and the SETI Institute -- are fundamentally
flawed. Some of the alternatives have existed for quite a long
time, starting with the seminal paper by Dyson (1960) and
elaborated in Dyson (1966) and \'Cirkovi\'c and Bradbury (2006).
What we can dub the Dysonian approach to SETI puts the emphasis
on the search for extraterrestrial technological manifestations
and artifacts. Even if they are not actively communicating with
us, that does not imply that we cannot detect their astro-
engineering activities. Unless advanced technological communities
have taken great lengths to hide or disguise their IR detection
signatures, the terrestrial observers should still be able to
observe them at those wavelengths and those should be
distinguishable from normal stellar spectra. In addition, other
bold unconventional studies like those on antimatter-burning
signatures (Harris 1986, 2002, Zubrin 1995), anomalous lines in
stellar spectra (Valdes and Freitas 1986), or recognizable
transits of artificial objects (Arnold 2005), seem to be
promising in ways conventional SETI is not. Search for
megaprojects such as Dyson Shells, Jupiter Brains or stellar
engines, are most likely to be successful in the entire spectrum
of SETI activities (Slysh 1985, Jugaku et al. 1995, Timofeev et
al. 2000, Jugaku and Nishimura 2003, Carrigan 2008).

All in all, considering the pace of the astrobiological
revolution, these issues are likely to be explored more and more
in years and decades to come. It is to be hoped that future
missions like TPF (Howard and Horowitz 2001), GAIA (Perryman et
al.\ 2001), or DARWIN (Cockell et al.\ 2009) will be able to offer
further quantitative inputs for the development of future, more
detailed numerical models of the astrobiological evolution of the
Milky Way (cf.\ Vukoti\'c and \'Cirkovi\'c 2008, Forgan 2009,
Cotta and Morales 2009). The overarching role played by
observation-selection effects in a large part of the relevant
hypothesis-space makes further research in this rather new field
mandatory from both points of view discussed above: research in
SETI {\bf and} research on the future of humanity. Resolving FP
is not a luxury; rather, it is one of the principal imperatives
if we wish our scientific worldview to have even a remote
prospect of completeness.

\vspace{0.5cm}

\noindent {\bf Acknowledgements.} The author thanks Stephen Webb,
Jelena Andreji\'c, Anders Sandberg, Branislav Vukoti\'c, Robert J.
Bradbury, Richard Cathcart, Irena Dikli\'c, Cl\'{e}ment Vidal,
George Dvorsky, J. Gaverick Matheny, Vjera Miovi\'c, Fred C.
Adams, Maja Bulatovi\'c, Alan Robertson, Nikola Milutinovi\'c,
Vesna Milo\v sevi\'c-Zdjelar, Samir Salim, Branislav K.
Nikoli\'c, Zoran Kne\v zevi\'c, Nikola Bo\v zi\'c, Mark A.
Walker, John Smart, Willian Lane Craig, Nick Bostrom, Slobodan
Popovi\'c, and Martin Beech for help in finding some of the
references or enjoyable discussions on the subject of this
review. A particular intellectual debt is owed to Zoran \v
Zivkovi\'c, who as a publisher, translator, critic and author of
fiction has, during the last three decades, done more than anyone
else to make me interested in the topic of extraterrestrial life
and intelligence. Special thanks are due to Du\v san Indji\'c and
Srdjan Samurovi\'c for kind help with the illustrations and
general support, as well as to Damian Veal for careful
proof-reading. This is also an opportunity to thank KoBSON
Consortium of Serbian libraries, which at last enabled overcoming
of the gap in obtaining the scientific literature during the
tragic 1990s. The author acknowledges support of the Ministry of
Science and Technological Development of the Republic of Serbia
through the project ON146012.

\vspace{0.5cm}

\section*{References}
\refe Ahouse, J. C. 1998, \textit{Biology and Philosophy\/} {\bf
13}, 359-391.

\refe All\`{e}gre, C. J., Manh\`{e}s, G., and G\"{o}pel, C. 1995,
{\it Geochim. Cosmochim. Acta\/} {\bf 59}, 1445-1456.

\refe Alm\'{a}r, I. 1992, {\it Acta Astronautica\/} {\bf 26},
253-256.

\refe Alvarez, L. W., Alvarez, W., Asaro, F. and Michel, H. V.
1980, {\it Science\/} {\bf 208}, 1095-1108.

\refe Ambrose, S. H. 1998, \textit{Journal of Human Evolution\/}
{\bf 34}, 623-651.

\refe Andrews, D. G. 2003, "Interstellar Transportation using
Today's Physics," AIAA Paper 2003-4691, report to 39th Joint
Propulsion Conference \& Exhibit.

\refe Annis, J. 1999a, \textit{J. Brit. Interplan. Soc.}
\textbf{52}, 19-22.

\refe Annis, J. 1999b, \textit{J. Brit.   Interplan. Soc.} {\bf
52}, 33-36.

\refe Arkhipov, A. V. 1996, {\it Observatory\/} {\bf 116},
175-176.

\refe Arkhipov, A. V. 1997, {\it Ap \& SS\/} {\bf 252}, 67-71.

\refe Arnold, L. F. A. 2005, {\it Astrophys. J.} {\bf 627},
534–539.

\refe Asghari, N. et al. 2004, {\it Astron. Astrophys.} {\bf
426}, 353-365.

\refe Bada, J. L. 2004, {\it Earth Planet. Sci. Lett.\/} {\bf
226}, 1-15.

\refe Badescu, V. 1995, {\it Acta Astronautica\/} {\bf 36}
135-138.

\refe Badescu, V. and Cathcart, R. B. 2000, {\it J. Brit.
Interplane. Soc.} {\bf 53}, 297-306.

\refe Badescu, V. and Cathcart, R. B. 2006, {\it Acta
Astronautica\/} {\bf 58} 119-129.

\refe Ball, J. A. 1973, {\it Icarus\/} {\bf 19}, 347-349.

\refe  Balashov, Yu. 1994, {\it Studies in History and Philosophy
of Science\/} {\bf 25B}, 933-958.

\refe Barkun, M. 2003, {\it A Culture of Conspiracy\/}
(University of California Press, Berkeley).

\refe Barrow, J. D. and Tipler, F. J. 1986, \textit{The Anthropic
Cosmological Principle\/} (Oxford  University Press, New York).

\refe Bauer, H. 1984, {\it Beyond Velikovsky: The History of a
Public Controversy\/} (University of Illinois Press, Urbana).

\refe Baxter, S. 2000, \textit{J. Br. Interplan. Soc.} {\bf 54},
210-216.

\refe Beaug\'e, C., Callegari, N., Ferraz-Mello, S., and
Michtchenko, T. A. 2005, in \textit{Dynamics of Populations of
Planetary Systems}, Proceedings of the IAU Colloquium No. 197, ed.
by Z. Kne\v zevi\'c and A. Milani (Cambridge University Press,
Cambridge), 3-18.

\refe Beech, M. 1990, \textit{Earth, Moon, and Planets\/} {\bf
49}, 177-186.

\refe Beech, M. 2008, \textit{Rejuvenating the Sun and Avoiding
Other Global Catastrophes\/} (Springer, New York).

\refe Benford, G. 1977, \textit{In the Ocean of Night\/} (The Dial
Press/James Wade, New York).

\refe Benford, G. 1983, \textit{Across the Sea of Suns\/} (Simon
\& Schuster, New York).

\refe Bjork, R. 2007, \textit{International Journal of
Astrobiology\/} {\bf 6}, 89-93.

\refe Bondi, H. and Gold, T. 1948, \textit{Mon. Not. R. astr.
Soc.} {\bf 108}, 252-270.

\refe Bostrom. N. 2000, {\it Futures\/} {\bf 35}, 759-764.

\refe Bostrom, N. 2002, \textit{Anthropic Bias: Observation
Selection Effects in Science and Philosophy\/} (Routledge, New
York).

\refe Bostrom, N. 2003, \textit{Philosophical Quarterly\/} {\bf
53}, 243-255.

\refe Bostrom, N. 2008 \textit{MIT Technology Review}, May/June
issue, 72-77

\refe Bostrom, N. and Cirkovic, M. M. (eds.) 2008, \textit{Global
Catastrophic Risks\/} (Oxford University Press, Oxford).

\refe Brillouin, L. 1962, {\it Science and Information Theory\/}
(Academic Press, New York).

\refe  Brin, G. D. 1983, \textit{Q. Jl. R. astr. Soc.}
\textbf{24}, 283-309.

\refe Browne, D. 2004, {\it Biology and Philosophy\/} {\bf 19},
633-653.

\refe Butterfield, H. 1962, \textit{The Origins of Modern
Science\/} (Collier, New York).

\refe Caplan, B. 2008, in \textit{Global Catastrophic Risks\/} by
N. Bostrom and M. M. \'Cirkovi\'c (eds.) (Oxford University Press,
Oxford, 2008), 498-513.

\refe Carter, B. 1983, {\it Philos. Trans. R. Soc. London A\/}
{\bf 310}, 347-363.

\refe Cavicchioli, R. 2002, {\it Astrobiology\/} {\bf 2}, 281-292.

\refe Chernavskii, D. S. 2000, {\it Physics--Uspekhi\/} {\bf 43},
151-176.

\refe Chyba, C. F. 1997, in \textit{Astronomical and Biochemical
Origins and the Search for Life in the Universe}, ed. by C. B.
Cosmovici, S. Bowyer, and D. Werthimer (Editrice Compositori,
Bologna), 157-164.

\refe Chyba, C. F. and Hand, K. P. 2005, \textit{Annual Review of
Astronomy \& Astrophysics\/} {\bf 43}, 31-74.

\refe Clark, R. 1968, \textit{JBS: The Life and Work of J. B. S.
Haldane\/} (Quality Book Club, London).

\refe Clarke, J. N. 1981, \textit{Icarus }\textbf{46}, 94-96.

\refe Clube, S. V. M. and Napier, W. M. 1984, \textit{Mon. Not. R.
Astron. Soc.} {\bf 211}, 953-968.

\refe Clube, S. V. M. and Napier, W. M. 1990, {\it The Cosmic
Winter\/} (Basil Blackwell Ltd, Oxford).

\refe \'Cirkovi\'c, M. M. 2004a, \textit{ J. Brit. Interplan.
Soc.} {\bf 57}, 53-59.

\refe \'Cirkovi\'c, M. M. 2004b, {\it Astrobiology\/} {\bf 4},
225-231.

\refe \'Cirkovi\'c, M. M. 2005, {\it J. Brit. Interplan. Soc.}
{\bf 58}, 62-70.

\refe \'Cirkovi\'c, M. M. 2006, \textit{Biology and Philosophy}
{\bf 21}, 369-379.

\refe \'Cirkovi\'c, M. M. 2007, \textit{International Journal of
Astrobiology\/} {\bf 6}, 325-329.

\refe \'Cirkovi\'c, M. M. 2008, {\it J. Brit. Interplan. Soc.}
{\bf 61}, 246-254.

\refe \'Cirkovi\'c, M. M. and Cathcart, R. B. 2004,  {\it J.
Brit. Interplan. Soc.} {\bf 57}, 209-215.

\refe \'Cirkovi\'c, M. M., Dragi\'cevi\'c, I., and
Beri\'c-Bjedov, T. 2005, {\it Serb. Astron. J.} {\bf 170}, 89-100.

\refe \'Cirkovi\'c, M. M. and Bradbury, R. J. 2006, \textit{New
Ast.} {\bf 11}, 628-639.

\refe \'Cirkovi\'c, M. M. and Vukoti\'c, B. 2008, \textit{Origin
of Life and Evolution of the Biosphere\/} {\bf 38}, 535-547.

\refe Cockell, C. S. et al. 2009, \textit{Astrobiology\/} {\bf
9}, 1-22.

\refe Cohen, J. and Stewart, I. 2002, {\it What Does a Martian
Look Like?\/} (John Wiley \& Sons, Hoboken, New Jersey).

\refe Conway Morris, S. 2003, \textit{Life's Solution: Inevitable
Humans in a Lonely Universe\/} (Cambridge University Press,
Cambridge).

\refe Cotta, C. and Morales, A. 2009, {\it J. Brit. Interplan.
Soc.} {\bf 62}, 82-88.

\refe Courtillot, V. 1999, \textit{Evolutionary Catastrophes\/}
(Cambridge University Press, Cambridge).

\refe Crick, F. H. C. and Orgel, L. E. 1973, \textit{Icarus\/}
{\bf 19}, 341-346.

\refe Criswell, D. R. 1985, in R. Finney and E.M. Jones (Eds),
\textit{Interstellar Migration and the Human Experience\/}
(University of California Press, Berkeley), 50-87.

\refe Crow, M. J. 1999, \textit{The Extraterrestrial Life Debate,
1750-1900\/} (Dover, Mineola, New York).

\refe Deardorff, J. W. 1986, \textit{Q. Jl. R. astr. Soc.}
\textbf{27}, 94-101.

\refe Deardorff, J. W. 1987, {\it J. Brit. Interplanet. Soc.}
{\bf 40}, 373-379.

\refe Dennett, D. C. 1995, \textit{Darwin's Dangerous Idea\/}
(Simon \& Schuster, New York, 1995).

\refe Des Marais, D. J. and Walter, M. R. 1999, \textit{Annu. Rev.
Ecol. Syst.} {\bf 30},    397-420.

\refe Dick, S. J., 1996, \textit{The Biological Universe: The
Twentieth Century Extraterrestrial Life Debate and the Limits of
Science \/} (Cambridge University Press, Cambridge).

\refe Dick, S. J. (ed.) 2000, \textit{Many Worlds\/} (Templeton
Foundation Press, Philadelphia).

\refe Dick, S. J., 2003,  \textit{Int. J. Astrobiology\/} {\bf
2}, 65-74.

\refe Drake, F. 1965, in \textit{Current Aspects of Exobiology},
ed. G. Mamikunian and M. H. Briggs (Pergamon, New York), 323-345.

\refe Duric, N. and Field, L. 2003, {\it Serb. Astron. J.} {\bf
167}, 1-10.

\refe Dyson, F. J. 1960, {\it Science\/} {\bf 131}, 1667-1668.

\refe Dyson, F. J., 1966, in Marshak, R.E. (ed), Perspectives in
Modern Physics, Interscience Publishers, New York, 641-655.

\refe Ehrenfreund, P. et al. 2002, {\it Rep. Prog. Phys.} {\bf
65}, 1427-1487.

\refe Eigen, M. 1992, \textit{Steps towards Life\/} (Oxford
University Press, Oxford).

\refe Erwin, D. H. 1993, {\it The Great Paleozoic Crisis: Life and
Death in the Permian\/} (Columbia University Press, New York).

\refe Fogg, M. J. 1987, \textit{Icarus\/} {\bf 69}, 370-384.

\refe Forgan, D. 2009, \textit{International Journal of
Astrobiology\/} {\bf 8}, in press.

\refe Freitas, R. A. Jr., 1985, {\it J. Brit. Interplanet. Soc.}
{\bf 38}, 106-112.

\refe Freitas, R. A., Jr. and Valdes, F., 1980,
\textit{Icarus\/} {\bf 42}, 442-447.

\refe Fry, I. 1995, \textit{Biology and Philosophy} {\bf 10},
389-417.

\refe Fry, I. 2000, \textit{The Emergence of Life on Earth\/}
(Rutgers University Press, New Brunswick).

\refe Galante, D. and Horvath, J. E. 2007, \textit{Int. J.
Astrobiol.} {\bf 6}, 19-26.

\refe Gehrels, N., Laird, C. M., Jackman, C. H., Cannizzo, J. K.,
Mattson, B. J., and Chen, W.    2003, \textit{Astrophys. J.} {\bf
585}, 1169-1176.

\refe Gies, D. R. and Helsel, J. W. 2005, {\it Astrophys. J.} {\bf
626}, 844-848.

\refe Gonzalez, G. 2005, \textit{Origin of Life and Evolution of
the Biosphere\/} {\bf 35}, 555-606.

\refe Gonzalez, G., Brownlee, D., and Ward, P. 2001,
\textit{Icarus\/} \textbf{152}, 185-200.

\refe Gould, S. J. 1987, {\it Time's Arrow, Time's Cycle\/}
(Harvard University Press, Cambridge).

\refe Greaves, J. S., Wyatt, M. C., Holland, W. S., and Dent, W.
R. F. 2004, \textit{Monthly Notices of the Royal Astronomical
Society\/} {\bf 351}, L54-L58.

\refe Grinspoon, D., 2003, \textit{Lonely Planets: The Natural
Philosophy of Alien Life\/} (HarperCollins, New York).

\refe Gros, C. 2005, \textit{J. Br. Interplan. Soc.} {\bf 58},
108-111.

\refe Haldane, J. B. S. 1972 [1927], \textit{Possible Worlds and
Other Essays\/} (Chatto and Windus, London).

\refe Hanslmeier, A. 2009, \textit{Habitability and Cosmic
Catastrophes\/} (Springer, Berlin).

\refe \refe Hanson, R. 1998, "The great filter - are we almost
past it?" preprint available at {\tt
http://hanson.gmu.edu/greatfilter.html}.

\refe Haqq-Misra, J. D. and Baum, S. D. 2009, \textit{J. Brit.
Interpl. Soc.} {\bf 62}, 47-51.

\refe Harris, M. J., 1986, \textit{Astrophys. Space Sci.} {\bf
123}, 297-303.

\refe Harris, M. J., 2002, \textit{J. Brit. Interpl. Soc.} {\bf
55}, 383-393.

\refe Hart, M. H. 1975, {\it Q. Jl. R. astr. Soc.} {\bf 16},
128-135.

\refe Hatcher, W. S. 1982, \textit{The Logical Foundations of
Mathematics\/} (Pergamon, London).

\refe Hawks, J. D. and Wolpoff, M. H. 2001, \textit{Evol. Int. J.
Org. Evol.} {\bf 55}, 1474-1485.

\refe Ho, D. and Monton, B. 2005, \textit{Analysis\/} {\bf 65},
42-45.

\refe Horner, J. and Jones, B. W. 2008, \textit{International
Journal of Astrobiology} {\bf 7}, 251-261.

\refe Horner, J. and Jones, B. W. 2009, \textit{International
Journal of Astrobiology}, in press.

\refe Howard, A. and Horowitz, P. 2001, \textit{Icarus\/} {\bf
150}, 163-167.

\refe Hoyle, F. and Wickramasinghe, N. C.  1981,
\textit{Evolution from Space\/} (J. M. Dent and Sons, London).

\refe Hoyle, F. and Wickramasinghe, N. C. 1999,
\textit{Astrophys. Space Sci.} {\bf 268} 89-102.

\refe Huggett, R. 1997, \textit{Catastrophism: Asteroids, Comets,
and Other Dynamic Events in Earth History\/} (Verso, London).

\refe Jablonski, D. 1986, \textit{Science\/} {\bf 231}, 129-133.

\refe Jaynes, J. 1990, {\it The Origin of Consciousness in the
Breakdown of the Bicameral Mind\/} (Houghton Mifflin, New York).

\refe Jones, E. M. 1976, \textit{Icarus\/} {\bf 28}, 421-422.

\refe Jones, E. M. 1981, \textit{Icarus\/} {\bf 46}, 328-336.

\refe Jugaku, J., Noguchi, K., and Nishimura, S. 1995, in
\textit{Progress in the Search for Extraterrestrial Life}, ed. by
G. Seth Shostak (ASP Conference Series, San Francisco), 181-185.

\refe Jugaku, J. and Nishimura, S. 2003, in \textit{Bioastronomy
2002: Life Among the Stars, Proceedings of IAU Symposium \# 213},
ed. by R. Norris and F. Stootman (ASP Conference Series, San
Francisco), 437-438.

\refe Kardashev, N. S. 1964, {\it Sov. Astron.} {\bf 8}, 217-220.

\refe Kinouchi, O. 2001, "Persistence solves Fermi paradox but
challenges SETI projects," preprint cond-mat/0112137.

\refe Korycansky, D. G., Laughlin, G., and Adams, F. C. 2001,
{\it Astrophys. Space Sci.} {\bf 275}, 349-366.

\refe Kragh, H. 1996, {\it Cosmology and Controversy\/} (Princeton
University Press, Princeton).

\refe Kragh, H. S. 2007, {\it Conceptions of Cosmos\/} (Oxford
University Press, Oxford).

\refe Kurzweil, R. 2005, \textit{The Singularity is Near\/}
(Duckworth, London).

\refe Kuhn, T. S. 1957, \textit{The Copernican Revolution\/}
(Harvard University Press, Cambridge).

\refe Lahav, N., Nir, S., and Elitzur, A. C. 2001, {\it Progress
in Biophysics \& Molecular Biology\/} {\bf 75}, 75-120.

\refe Landis, G. A. 1998, \textit{J. Brit. Interplan. Soc.} {\bf
51}, 163-166.

\refe L\'eger, A., Selsis, F., Sotin, C., Guillot, T., Despois,
D., Mawet, D., Ollivier, M., Lab\'eque, A., Valette, C., Brachet,
F., Chazelas, B., and Lammer, H. 2004, {\it Icarus\/} {\bf 169},
499.

\refe Leitch, E. M. and Vasisht, G. 1998,  \textit{New Ast.} {\bf
3}, 51-56.

\refe Lem, S. 1977, \textit{Summa Technologiae\/} (Nolit,
Belgrade, in Serbian).

\refe Lem, S. 1984, \textit{His Master's Voice\/} (Harvest Books,
Fort Washington).

\refe Lem, S. 1987, \textit{Fiasco\/} (Harcourt, New York).

\refe Lineweaver, C. H. 2001, \textit{Icarus} \textbf{151},
307-313.

\refe Lineweaver, C. H. and Davis, T. M. 2002, {\it
Astrobiology\/} {\bf 2}, 293-304.

\refe Lineweaver, C. H., Fenner, Y., and Gibson, B. K. 2004,
{\it Science\/} {\bf 303}, 59-62.

\refe Lipunov, V. M. 1997, \textit{Astrophys. Space Sci.} {\bf
252}, 73-81.

\refe Livio, M. 1999, \textit{Astrophys. J.} {\bf 511}, 429-431.

\refe Lovecraft, H. P. 2005 [1931], \textit{At the Mountains of
Madness: The Definitive Edition\/} (Random House, New York).

\refe Lytkin, V., Finney, B., and Alepko, L. 1995, \textit{Q. J.
R. astr. Soc.} {\bf 36}, 369-376.

\refe Maher, K. A. and Stevenson, D. J. 1988, \textit{Nature\/}
{\bf 331}, 612-614.

\refe Mayr, E., 1993, \textit{Science\/} {\bf 259}, 1522-1523.

\refe McInnes, C. R. 2002, {\it Astrophys. Space Sci.} {\bf 282},
765-772.

\refe Melott, A. L.\ et al.\ 2004, {\it International Journal of
Astrobiology\/} {\bf 3}, 55-61.

\refe M\'{e}sz\'{a}ros, P. 2002, {\it Annual Review of Astronomy
and Astrophysics\/} {\bf 40}, 137-169.

\refe Michaud, M. A. G. 2007, \textit{Contact with Alien
Civilizations\/} (Springer, New York).

\refe Mojzsis, S. J., Arrhenius, G., McKeegan, K. D., Harrison, T.
M., Nutman, A. P., and Friend, C. R. L. 1996, {\it Nature\/} {\bf
384}, 55-59.

\refe Newman, W. I. and Sagan, C. 1981, \textit{Icarus\/} {\bf
46}, 293-327.

\refe Noble, M., Musielak, Z. E., Cuntz, M. 2002, {\it Astrophys.
J.} {\bf 572}, 1024-1030.

\refe N\o rretranders, T. 1999, \textit{The User Illusion:
Cutting Consciousness Down to Size\/} (Penguin, New York).

\refe Norris, R. P. 2000, in \textit{When SETI Succeeds: The
impact of high-information Contact}, ed. A. Tough (Foundation for
the Future, Washington DC), 103-105.

\refe Olum, K. D. 2004, \textit{Analysis\/} {\bf 64}, 1-8.

\refe Palmer, T. 2003, {\it Perilous Planet Earth: Catastrophes
and Catastrophism through the Ages\/} (Cambridge University Press,
Cambridge).

\refe Parkinson, B. 2004, \textit{J. Br. Interplan. Soc.} {\bf
57}, 60-66.

\refe Pavlov, A. A., Toon, O. B., Pavlov, A. K., Bally, J., and
Pollard, D. 2005, \textit{Geophys. Res. Lett.} {\bf 32}, L03705
(1-4).

\refe Pe\~na-Cabrera, G. V. Y., and Durand-Manterola, H. J. 2004,
{\it Adv. Space Res.} {\bf 33}, 114-117.

\refe Penrose, R. 1989, \textit{The Emperor's New Mind\/} (Oxford
University Press, Oxford).

\refe Rampino, M. R. 2002, \textit{Icarus\/} {\bf 156}, 562-569.

\refe Raup, D. M. 1991, {\it Extinction: Bad Genes or Bad Luck?\/}
(W. W. Norton, New York).

\refe Raup, D. M. 1992, \textit{Acta Astronautica\/} {\bf 26},
257-261.

\refe Raup, D. M. and Valentine, J. W. 1983, \textit{Proc. Natl.
Acad. Sci. USA\/} {\bf 80}, 2981-2984.

\refe Reynolds, A. 2002 \textit{Revelation Space\/} (Gollancz,
London).

\refe Reynolds, A. 2004, \textit{Century Rain\/} (Gollancz,
London).

\refe Rose, C. and Wright, G. 2004, {\it Nature\/} {\bf 431},
47-49.

\refe Rummel, J. D. 2001, \textit{Proceedings of the National
Academy of Science\/} {\bf 98}, 2128-2131.

\refe Russell, D. A. 1983, \textit{Adv. Space Res.} {\bf 3},
95-103.

\refe Saberhagen, F. 1998, \textit{Berserkers: The Beginning\/}
(Baen, Riverdale, New York).

\refe Sagan, C., Ed. 1973, \textit{Communication with
Extraterrestrial Intelligence\/} (MIT Press,  Cambridge).

\refe Sagan, C., and Walker, R. G., 1966, \textit{Astrophys. J.}
{\bf 144}, 1216-1218.

\refe Sagan, C. and Newman, W. I. 1983, \textit{Q. Jl. R. astr.
Soc.} {\bf 24}, 113-121.

\refe Sandberg, A. 2000, \textit{Journal of Evolution and
Technology\/} {\bf 5} (at \\ {\tt
http://transhumanist.com/volume5/Brains2.pdf}).

\refe Scalo, J. and Wheeler, J. C. 2002, \textit{Astrophys. J.}
\textbf{566}, 723-737.

\refe Schaller, R. R. 1997,  {\it IEEE Spectrum}, June 1997,
53-59.

\refe Schroeder, K. 2002, {\it Permanence\/} (Tor Books, New
York).

\refe Shaviv, N. J. 2002,  {\it New Ast.} {\bf 8}, 39-77.

\refe Shklovskii, I. S. and Sagan, C. 1966, \textit{Intelligent
Life in the Universe\/} (Holden-Day, San Francisco).

\refe Simpson, G. G., 1964, \textit{Science\/} {\bf 143}, 769-775.

\refe Slysh, V. I. 1985, in \textit{The Search  for
Extraterrestrial Life: Recent Developments}, ed. by M. D.
Papagiannis (IAU, Reidel Publishing Co., Dordrecht), 315-319.

\refe Smart, J. J. C. 2004, \textit{Stud. Hist. Phil. Biol. \&
Biomed. Sci.} {\bf 35}, 237-247.

\refe Smart, J. 2007, "Answering the Fermi Paradox: Exploring the
Mechanisms of Universal Transcension," (preprint at \\ {\tt
http://accelerating.org/articles/answeringfermiparadox.html.})

\refe Tadross, A. L. 2003, \textit{New Astronomy }\textbf{8}, 737.

\refe Tarter, J. 2001, \textit{Annu. Rev. Astron.  Astrophys.}
{\bf 39}, 511-548.

\refe Terry, K. D. and Tucker, W. H. 1968, {\it Science\/} {\bf
159}, 421-423.

\refe Thomas, B. C., Jackman, C. H., Melott, A. L., Laird, C. M.,
Stolarski, R. S., Gehrels, N., Cannizzo, J. K., and Hogan, D. P.
2005, \textit{Astrophys. J.} {\bf 622}, L153-L156.

\refe Thomas, B. C., Melott, A. L., Fields, B. D., and
Anthony-Twarog, B. J. 2008, \textit{Astrobiology\/} {\bf 8}, 9-16.

\refe Thorsett, S. E. 1995, \textit{Astrophys. J.} \textbf{444},
L53-L55.

\refe Timofeev, M. Yu., Kardashev, N. S., and Promyslov, V. G.
2000, \textit{Acta Astronautica\/} {\bf 46}, 655-659.

\refe Tipler, F. J. 1980, \textit{Q. Jl. R. astr. Soc.} {\bf 21},
267-281.

\refe Tipler, F. J. 1981, \textit{Q. Jl. R. astr. Soc.} {\bf 22},
133-145.

\refe Valdes, F. and Freitas, R. A. Jr. 1986, {\it Icarus\/} {\bf
65}, 152$-$157.

\refe Vinge, V. 1986, \textit{Marooned in Realtime\/} (St.
Martin's Press, New York).

\refe Vinge, V. 1991, {\it A Fire upon the Deep\/} (Millenium,
London).

\refe Vinge, V. 1993, "The Coming Technological Singularity" in
\textit{Vision-21: Interdisciplinary Science \& Engineering in the
Era of CyberSpace}, proceedings of a Symposium held at NASA Lewis
Research Center (NASA Conference Publication CP-10129).

\refe von Hoerner, S. 1978, \textit{Die Naturwissenschaften\/}
{\bf 65}, 553-557.

\refe Vukoti\'c, B. and \'Cirkovi\'c, M. M. 2007,  \textit{Serb.
Astron. J.} {\bf 175}, 45-50.

\refe Vukoti\'c, B. and \'Cirkovi\'c, M. M. 2008, \textit{Serb.
Astron. J.} {\bf 176}, 71-79.

\refe Wallace, A. R. 1903, \textit{Man's Place in the Universe\/}
(McClure, Phillips \& Co., New York).

\refe Ward, P. D. and Brownlee, D. 2000, \textit{Rare Earth: Why
Complex Life Is Uncommon in the Universe }(Springer, New York).

\refe Webb, S. 2002, \textit{Where is Everybody? Fifty Solutions
to the Fermi's Paradox\/} (Copernicus, New York).

\refe Wesson, P. S. 1990, \textit{Q. Jl. R. astr. Soc.} {\bf
31}, 161-170.

\refe Wesson, P. S., Valle, K., and Stabell, R. 1987, {\it
Astrophys. J.} {\bf 317}, 601-606.

\refe Williams, D. M., Kasting, J. F., and Wade, R. A. 1997, {\it
Nature\/} {\bf 385}, 234-236.

\refe Wilson, P. A. 1994, {\it Brit. J. Phil. Sci.} {\bf 45},
241-253.

\refe Woosley, S. E. and Bloom, J. S. 2006, {\it Annual Review of
Astronomy and Astrophysics\/} {\bf 44}, 507-556.

\refe Zubrin, R. 1995, in \textit{Progress in the Search for
Extraterrestrial Life}, ed. by G. Seth Shostak (ASP Conference
Series, San Francisco), 487-496.

\refe Zuckerman, B. 1985, {\it Q. Jl. R. astr. Soc.} {\bf 26},
56-59.
\end{document}